\documentclass[preprint]{aastex}
\usepackage{emulateapj5,apjfonts}

\newcommand  \kms    {\ifmmode {\rm km\,s}^{-1} \else km\,s$^{-1}$\fi}
\newcommand  \cc     {\hbox{cm$^{-3}$}}

\newcommand  \ergcms {\ifmmode {\rm ergs\,cm}^{-2}\,{\rm s}^{-1}
               \else ergs\,cm$^{-2}$\,s$^{-1}$\fi}

\slugcomment{Accepted by The Astrophysical Journal}

\shorttitle{X-RAY ABSORBING AND EMITTING OUTFLOW IN NGC\,3783}

\shortauthors{KASPI ET AL.}

\begin{document}

\title{High-Resolution X-ray Spectroscopy and Modeling of
the Absorbing and Emitting Outflow in NGC\,3783}
\author{
Shai Kaspi,\altaffilmark{1} 
W. N. Brandt,\altaffilmark{1} 
Hagai Netzer,\altaffilmark{2} 
Ian M. George,\altaffilmark{3,4}
George Chartas,\altaffilmark{1}
Ehud Behar,\altaffilmark{5}
Rita M. Sambruna,\altaffilmark{1,6}
Gordon P. Garmire,\altaffilmark{1} 
and
John A. Nousek\altaffilmark{1}
}
\altaffiltext{1}{Department of Astronomy and Astrophysics, 525 Davey
Laboratory, The Pennsylvania State University, University Park, PA 16802.}
\altaffiltext{2}{School of Physics and Astronomy, Raymond and Beverly
Sackler Faculty of Exact Sciences, Tel-Aviv University, Tel-Aviv 69978, Israel.}
\altaffiltext{3}{Laboratory for High Energy Astrophysics, Code 662,
NASA/Goddard Space Flight Center, Greenbelt, MD 20771.}
\altaffiltext{4}{Joint Center for Astrophysics, University of Maryland,
Baltimore County, 1000 Hilltop Circle, Baltimore, MD 21250.}
\altaffiltext{5}{Columbia Astrophysics Laboratory, Columbia University,
New York, NY 10027-5247.}
\altaffiltext{6}{Department of Physics \& Astronomy and School of
Computational Sciences, George Mason University, 4400 University Dr.
M/S 3F3, Fairfax, VA 22030-4444.}

\begin{abstract}
The high-resolution X-ray spectrum of NGC\,3783 shows several dozen
absorption lines and a few emission lines from the H-like and He-like
ions of O, Ne, Mg, Si, and S as well as from \ion{Fe}{17}--\ion{Fe}{23}
L-shell transitions. We have reanalyzed the {\it
Chandra} HETGS spectrum using better flux and wavelength calibrations
along with more robust methods. Combining several lines from each
element, we clearly demonstrate the existence of the absorption lines
and determine they are blueshifted relative to the systemic velocity by
$-610\pm 130$ \kms. We find the Ne absorption lines in the High Energy
Grating spectrum to be resolved with ${\rm FWHM}=840^{+490}_{-360}$
km\,s$^{-1}$; no other lines are resolved.  The emission lines are
consistent with being at the systemic velocity. We have used regions in
the spectrum where no lines are expected to determine the X-ray
continuum, and we model the absorption and emission lines using
photoionized-plasma calculations. The model consists of two absorption
components, with different covering factors, which have an order of
magnitude difference in their ionization parameters. The two components
are spherically outflowing from the AGN and thus contribute to both the
absorption and the emission via P Cygni profiles. The model also
clearly requires \ion{O}{7} and \ion{O}{8} absorption edges.  The
low-ionization component of our model can plausibly produce UV
absorption lines with equivalent widths consistent with those observed
from NGC\,3783. However, we note that this result is highly sensitive
to the unobservable UV-to-X-ray continuum, and the available UV and
X-ray observations cannot firmly establish the relationship between the
UV and X-ray absorbers. We find good agreement between the {\it
Chandra} spectrum and simultaneous {\it ASCA} and {\it RXTE}
observations. The 1 keV deficit previously found when modeling {\it
ASCA} data probably arises from iron L-shell absorption lines not
included in previous models. We also set an upper limit on the FWHM of
the narrow Fe~K$\alpha$ emission line of 3250~\kms. This is consistent
with this line originating outside the broad line region, possibly from
a torus.
\end{abstract}


\section{Introduction}

Active Galactic Nuclei (AGNs) often show evidence for deep absorption
features from 0.7--1.5~keV which have been typically attributed to
\ion{O}{7} (739~eV) and \ion{O}{8} (871~eV) edges. The ionized gas
component creating these features is usually referred to as a `warm
absorber' and is seen in many Seyfert~1 galaxies (e.g., Reynolds
1997; George et~al. 1998b) and some quasars (e.g., Halpern
1984; Brandt et~al. 1997). The fact that these absorption features are
seen in $\gtrsim$70\% of Seyfert~1s implies that the ionized gas covers
a substantial fraction of the central X-ray source. The radial location
of the warm absorber is still not well constrained. Several Seyfert~1s
have shown warm absorber edge variability on timescales of hours to
weeks (e.g., Fabian et~al. 1994; Otani et~al. 1996; George et al.
1998a, 1998b), suggesting that they have some ionized gas lying at
distances characteristic of the Broad Line Region (BLR). Warm absorbers
are also a subject of intense theoretical investigation (e.g., Netzer
1993, 1996; Krolik \& Kriss 1995; Reynolds \& Fabian 1995; Nicastro,
Fiore, \& Matt 1999). They are expected to be emitters of X-ray lines
such as \ion{O}{7}~(568~eV), \ion{O}{8}~(653~eV), and
\ion{Ne}{9}~(915~eV). They are also expected to produce X-ray
absorption lines with significant equivalent widths (EWs).

% \begin{deluxetable}{ccccc}
% \tablecolumns{5}
% \tablewidth{0pt}
\begin{table*}
\footnotesize    % better fit the Journal font size.
\caption{Observation Log of NGC\,3783
\label{observations}}
\begin{center}
\begin{tabular}{ccccc}
\hline
\hline
{Observatory} &
{Sequence Number} &
{UT start} &
{UT end} &
{Time\tablenotemark{a}} \\
{(1)} &
{(2)} &
{(3)} &
{(4)} &
{(5)} \\
\hline
{\it Chandra}&700045\phn\phn & 2000 Jan 20, 23:33 & 2000 Jan 21, 16:20 & 56.0 \\
{\it ASCA}    & 77034000 & 2000 Jan 12, 06:23  & 2000 Jan 12, 21:21 & 15.6 \\
{\it ASCA}    & 77034010 & 2000 Jan 16, 21:00  & 2000 Jan 17, 13:05 & 17.6 \\
{\it ASCA}    & 77034020 & 2000 Jan 21, 04:02  & 2000 Jan 21, 18:27 & \phn4.1 \\
{\it RXTE}    & 30227\phn\phn\phn & 2000 Jan 21, 04:37  & 2000 Jan 21, 16:56 & 19.6 \\
\hline
\end{tabular}
\vskip 2pt
\parbox{4.0in}{ % use this to define the width of the notes under the table
\small\baselineskip 9pt
\footnotesize
\indent
$\rm ^a${Sum of good time intervals in units of ks.
For {\it ASCA} the time is the mean from the four detectors.}
}
\end{center}
\end{table*}
\normalsize

The bright Seyfert~1 galaxy NGC\,3783 ($V\approx13.5$ mag) has one of
the strongest warm absorbers known. Its X-ray spectrum and ionized
absorption have been modeled using data from {\em ROSAT} (Turner et
al. 1993) and {\em ASCA} (e.g., George et al. 1998a). The 2--10 keV
spectrum is fit by a power law with photon index $\Gamma \approx
1.7$--1.8. The 2--10 keV flux varies in the range
$\approx(4$--$9)\times 10^{-11}$~ergs\,cm$^{-2}$\,s$^{-1}$, and its
mean X-ray luminosity is $\approx 3\times 10^{43}$~ergs\,s$^{-1}$ (for
$H_0=50$~km\,s$^{-1}$\,Mpc$^{-1}$). Modeling the apparent \ion{O}{7}
and \ion{O}{8} edges indicates a column density of ionized gas of
$\approx 2 \times 10^{22}$~cm$^{-2}$. The {\it ASCA} spectra of
NGC\,3783 show excess flux around 600~eV, and this has been interpreted
as due to emission lines from the warm absorber, particularly
\ion{O}{7}~(568~eV) (e.g., George, Turner, \& Netzer 1995; George
et~al. 1998b). The {\it ASCA} spectra also show a deficit at $\approx
1$ keV compared with the expectations from a single-zone ionized
absorber model. George et al. (1998a) discuss several possible
explanations for this and favor the possibility of having two or more
zones of photoionized material. X-ray observations have also revealed
changes in the warm absorber of NGC\,3783 over time, but it has not
been possible to determine whether this variability is primarily due to
changes in the warm absorber's column density or ionization parameter
(see George et~al. 1998a for detailed discussion).

UV spectra of NGC\,3783 show intrinsic absorption features due to
\ion{C}{4}, \ion{N}{5} and \ion{H}{1} (e.g., Maran et al. 1996; Shields
\& Hamann 1997; Crenshaw et~al. 1999). Currently there are three known
absorption systems in the UV at radial velocities of approximately
$-$560, $-$800, and $-$1400~\kms\ (blueshifted) relative to the optical
redshift [throughout this paper we use a redshift of $0.00976\pm
0.000093$ ($\pm$28 \kms ) cited by de Vaucouleurs et al. (1991) which
was used also by Crenshaw et al. (1999)]. The strength of the
absorption is found to be variable. On 1992 July 27 and 1994 January 16
a \ion{C}{4} line was present at a radial velocity of $\approx -560$
km\,s$^{-1}$ while no such absorption was seen on 1993 February 5.
However, an observation on 1993 February~21 revealed \ion{N}{5}
absorption at $\approx -560$ km\,s$^{-1}$. A further observation from
1995 April 11 showed \ion{C}{4} absorption lines at velocities of
$\approx -560$ and $-$1400 \kms , and a recent observation with {\it
HST} STIS (2000 February 27; Crenshaw, Kramer, \& Ruiz 2000) revealed a
new third absorption component of \ion{C}{4} and \ion{N}{5} at $\approx
-800$~km\,s$^{-1}$. In addition the component at $\approx
-1400$~km\,s$^{-1}$ was much stronger and is present also in
\ion{Si}{4}, indicating a lower ionization state compared to the other
components. The \ion{C}{4} absorption lines are imprinted on the
\ion{C}{4} emission lines; if the X-ray and UV absorbers are connected,
this suggests that the warm absorber is located at a radius larger than
that of the BLR emitting the \ion{C}{4} ($\approx 4$ light days;
Reichert et al. 1994).

In Kaspi et al. (2000; hereafter Paper~I) we presented first results
from a high-resolution X-ray spectrum of NGC\,3783 obtained by {\em
Chandra}. We identified the main absorption and emission lines and used
them to construct a simple model of the absorbing gas in this AGN. In
this Paper we present a more detailed analysis combined with
simultaneous observations from {\em ASCA} and {\em RXTE}.

In the following sections we describe the observations and data
reduction from the three missions (\S~2), we describe time variability
(\S~3), and we model the X-ray continuum emission (\S~4). In \S~5 we
use calculations for photoionized plasma to model and discuss in detail
the high-resolution X-ray spectrum of NGC~3783. We discuss the
connection between the UV and X-ray absorbers in \S~6, and the narrow
and broad Fe K$\alpha$ emission lines are discussed in \S~7.

\section{Observations and Data Reduction}

A log of the observations from the satellites is presented in
Table~\ref{observations}. Below we detail issues concerning each of the observations.

\subsection{ {\it Chandra}}
\label{chandra_obs}

NGC\,3783 was observed using the High Energy Transmission Grating
Spectrometer (HETGS; C. R. Canizares et al., in preparation) on the
{\em Chandra X-ray Observatory\/}\footnote{See {\em The Chandra
Proposers' Observatory Guide} at http://asc.harvard.edu/udocs/docs/}
with the Advanced CCD Imaging Spectrometer (ACIS; G. P. Garmire et al.,
in preparation) as the detector. Brief descriptions of the observation
and data reduction are given in Paper~I. Here we repeat these with
additional details and emphasis on some changes.

The observation was continuous with a total integration time of 56~ks.
For this paper we re-reduced the data in 2000 July using the most
up-to-date {\em Chandra} Interactive Analysis of Observations ({\sc
ciao}) software (Version~1.1.4) and its calibration data. The resulting
spectra are essentially the same as those obtained in Paper~I with a
few minor deviations which are well within statistical errors.

The HETGS produces a zeroth order X-ray spectrum at the aim point on
the CCD and higher order spectra which have much higher spectral
resolution along the ACIS-S array. The higher order spectra are from
two grating assemblies, the High Energy Grating (HEG) and Medium Energy
Grating (MEG). Both positive and negative orders are imaged by the
ACIS-S array. Order overlaps are discriminated by the intrinsic energy
resolution of ACIS.

During this observation one of the six ACIS-S CCDs (S0) was shut off
due to problems with one of the instrument's front end processors. This
resulted in somewhat reduced wavelength coverage in the MEG and HEG
negative order spectra. However, at the relevant wavelengths
($>26.2$~\AA\ and $>13.1$~\AA\ in $-1$st order, respectively) few
counts are expected due to the Galactic column density ($N_{\rm
H}=8.7\times 10^{20}$~cm$^{-2}$; adopted from Alloin et al. 1995 and
consistent with Stark et al. 1992) and the low effective area of the
HETGS. Part of this wavelength region is covered (in the +1st order) by
the S5 CCD, and thus shutting off the S0 CCD resulted in the minimum
impact on the observation.

The zeroth-order spectrum of NGC~3783 shows substantial photon pileup.
Pileup results when two or more photons are detected as single event;
this distorts the energy spectrum and causes an underestimate of the
count rate.  Using {\sc marx} Version~3.0 (Wise et al. 1999) simulation
we established that this pileup produced a factor of $\approx 5.7$ loss
of events within $2\farcs5$. This effect could not be corrected, and we
did not extract spectral information from the zeroth-order image in
this study. Due to a problem in the aspect solution for {\it Chandra}
data processed before 2000 February 14, the image has an offset of
about 8\arcsec\ from the real celestial coordinates. We used {\sc ciao}
to correct for this positional offset, and the positions below should
be correct to within $\approx 1\arcsec$. The central pixel of the AGN
in the {\it Chandra} 0.5--8~keV image is at $\alpha_{2000}=$~11$^{\rm
h}$39$^{\rm m}$01$\fs$7,
$\delta_{2000}=$~$-37\arcdeg$44$\arcmin$19$\farcs$0. This coincides (to
within the uncertainties) with the AGN position reported in Ulvestad \&
Wilson (1984); our position is offset from theirs by 0$\farcs$4 to the
North-West. Comparing the point spread function (PSF) wings of the
NGC\,3783 image to the PSF of a point source (the HETGS observation of
Capella), we find very good agreement. Thus, we find no evidence for an
extended circumnuclear component in NGC\,3783. The only additional
point source detected that is coincident with the optical disk of
NGC\,3783 is the very faint object we reported on in Paper~I.

The first-order spectra from the MEG and HEG have signal-to-noise
ratios (S/Ns) of $\approx 5$ and $\approx 2.5$, respectively (at around
7~\AA ) when using the default {\sc ciao} bins of 0.005~\AA\ for the
MEG and 0.0025~\AA\ for the HEG. The higher-order spectra had only a
few ($\lesssim 3$) photons per bin and will not be discussed further
here. To obtain a high S/N first-order spectrum we binned each of the
MEG and HEG $\pm$1st order spectra into 0.01~\AA\ bins (which is still
about half the MEG resolution). The uncertainties in the binned counts
were computed using Gehrels (1986). Each of the four spectra was flux
calibrated using Ancillary Response Files produced by {\sc ciao}. The
uncertainty in the flux calibration is currently estimated to be
$\la$~30\%, $\la$~20\%, and $\la$~10\% in the 0.5--0.8, 0.8--1.5, and
1.5--6~keV bands, respectively (H. L. Marshall 2000, private
communication).\footnote{See
http://space.mit.edu/ASC/calib/hetgcal.html} We also corrected the
spectra for Galactic absorption and the cosmological redshift. The
background level of the X-ray spectrum is typically 0 or 1 counts per
dispersion bin, well within the Poissonian error of the spectrum. This
background is largely a result of cosmic rays which strike the CCDs
during the observation, and there are no obvious systematic effects in
it. Hence, no background subtraction was applied to the spectra
presented below.

During the analysis of these data the HETGS instrument team announced a
systematic wavelength calibration error that exists in standard
processed data (H. L. Marshall 2000, private communication).$^8$ This
error is traced to a small reduction in the ACIS-S pixel size due to
thermal contraction. The systematic error causes the data processed by
{\sc ciao} to have wavelengths larger by $\approx 0.054$\% than their
true values. Since there is no remedy for this effect in the current
version of {\sc ciao}, we manually corrected our final spectra by this
systematic factor. All the results reported in this paper take into
account this correction.

\begin{figure*}
\centerline{\includegraphics[width=18.4cm]{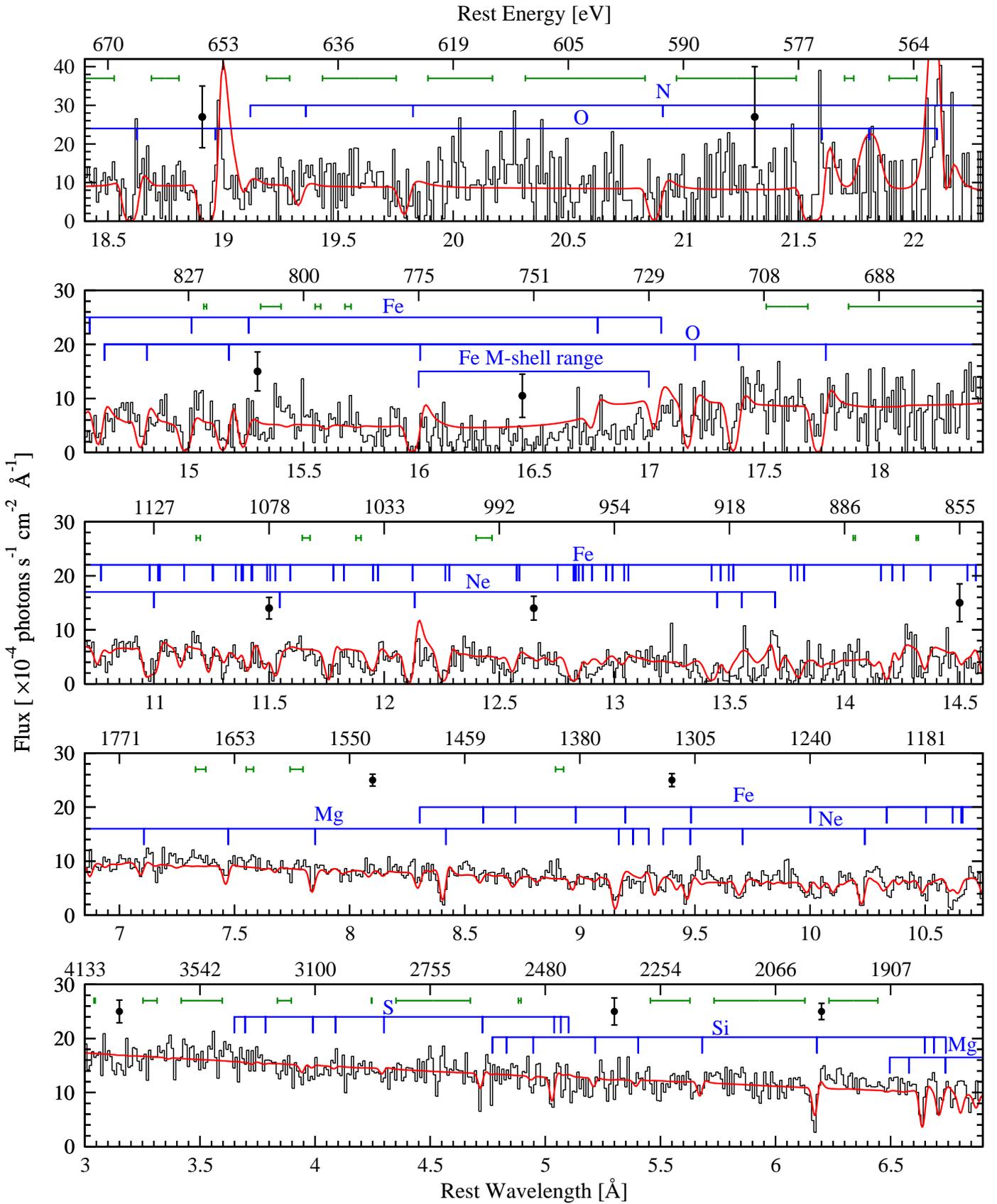}}
\caption{Combined MEG and HEG first-order spectrum of NGC\,3783 (black
line). Data are binned into 0.01~\AA\ bins and are not smoothed. The
spectrum has been corrected for Galactic absorption and the
cosmological redshift. Dots with error bars show the typical $\pm
1\sigma$ statistical errors at various wavelengths. The two-component
photoionization model discussed in \S~\ref{two_component} is
overplotted (red line) to emphasize its good agreement with the data.
The H-like and He-like lines of N, O, Ne, Mg, Si, and S, as well as the
strongest (EW$>$5 m\AA ) Fe lines contributing to the model are marked
with blue. The line-free zones are marked as horizontal green ranges.
\label{megspec} }
\end{figure*}

Since the $+1$st and $-1$st orders in each of the MEG and HEG spectra
are in excellent agreement (both in flux and wavelength), we averaged
them using a $1/\sigma^2$ weighted mean to produce mean MEG and HEG
spectra. After checking for consistency between the mean MEG and HEG
spectra, we also averaged these two spectra at wavelengths below
13\,\AA\ (above this wavelength there are no significant counts in the
HEG spectrum). The final spectrum is presented in
Figure~\ref{megspec}. The total number of counts in the spectrum
is 72900.

We note a small ($\approx 20$\%) systematic difference in flux between
the MEG $+1$st and $-1$st orders around 2.6--2.9 \AA\ where a chip gap
is present (at all other chip gaps there is good agreement in flux).
However, the effects of the chip-gaps on our combined spectrum are
small because of the weighted-mean method we used to derive it: since
the numbers of counts in the regions of chip gaps are relatively small
(and thus the associated statistical uncertainties are relatively
large), these regions receive less weight.

\subsection{{\it ASCA}}
\label{obsasca}

We observed NGC\,3783 with {\em ASCA} (Tanaka, Inoue, \& Holt 1994) in
3 epochs during 2000 January (see Table~\ref{observations}). The
multiple observations were designed to provide the time history of the
variable ionizing continuum in order to understand better the X-ray
emission-line strengths and the state of the object during the {\it
Chandra} observation. During each {\em ASCA} observation the two
Solid-state Imaging Spectrometers (known as SIS0 and SIS1;
$\approx$\,0.4--10~keV) and the two Gas Imaging Spectrometers (known as
GIS2 and GIS3; $\approx$\,0.8--10~keV) were operated. We restrict our
analysis of the data from the SIS detectors to that collected in {\sc
faint} mode. The data reduction and screening were performed using
{\sc ftools} (Version~5.0) as described in detail in George et al.
(1998a). The total exposure times are listed in
Table~\ref{observations}.

During the third observation, which was carried out simultaneously with
the {\it Chandra} and {\em RXTE} observations, there was a severe
disruption in the data down-link to a Deep Space Network station (K.
Mukai 2000, private communication). About 75\% of the data from this
observation were lost unrecoverably due to this failure.

\subsection{{\it RXTE}}

An {\em RXTE} (Bradt, Rothschild, \& Swank 1993) observation was
carried out simultaneously with the {\em Chandra} observation (see
Table~\ref{observations}). Here we present the results from the
large-area (0.7 m$^2$) Proportional Counter Array (PCA). The PCA
consists of 5 proportional counter units (PCUs) of which only 3 were
operational at the time of our observation (units 0, 2, and 3). All PCA
count rates in this paper refer to the total counts from these 3 PCUs.
We include only data from the upper detection layer as this layer
provides the highest S/N for photons in the energy range 2--20~keV.

We reduced the data in the standard way following the {\it RXTE} Cook
Book.\footnote{See
http://rxte.gsfc.nasa.gov/docs/xte/recipes/cook\_book.html} We used
standard ``good time interval'' criteria to select data with the lowest
background. We discarded data obtained when the Earth elevation angle
was less than 10$\degr$, when the pointing offset from the source was
$>0.02\degr$, or when there was significant electron contamination
({\sc electron0}\,{$>$}\,0.1). We also discarded data obtained during
passages through the South Atlantic Anomaly or up to 30 minutes after
the peaks of such passages. The PCA is a non-imaging device with a
field of view of FWHM~$\sim 1\degr$, and the background we subtracted
was calculated from a model for a faint source (using the {\sc ftools}
routine {\sc pcabackest} Version 2.1a).

\section{Time Variability}

In order to compare the state of the X-ray source in NGC\,3783 during
the {\it Chandra} observation to its state during past years, we use
the {\it ASCA} observations. For comparison purposes we followed the
variability analysis described in George et al. (1998a). We constructed
light curves for different energy ranges during each observation. The
light curves from SIS0 and \ SIS1 \ and \ from \ GIS2 \ and \ GIS3
\ were \ added \ in \ each \ case.

%\begin{figure}
\centerline{\includegraphics[width=8.5cm]{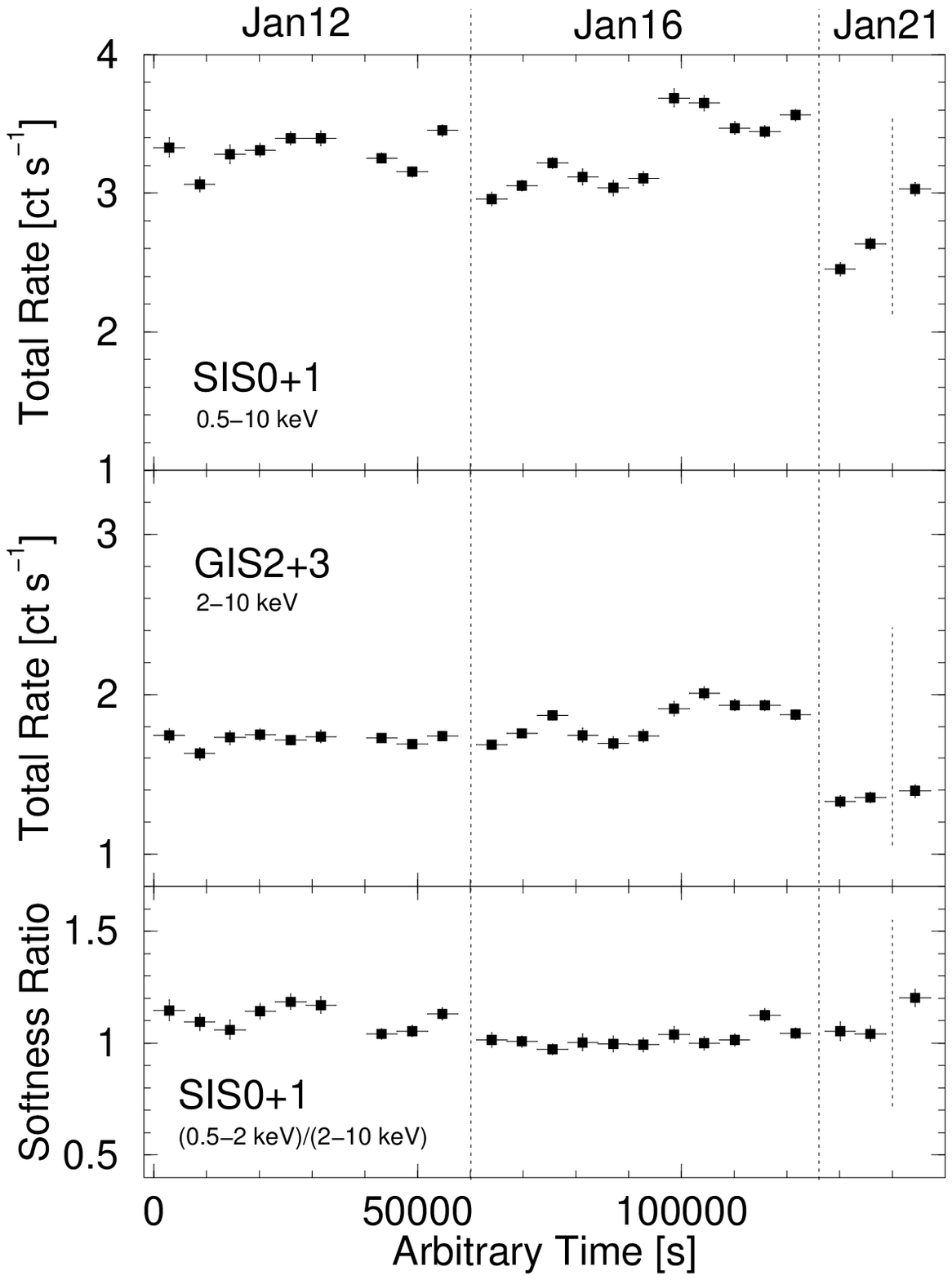}}
\figcaption{Light curves for the three {\em ASCA} observations of
NGC\,3783 carried out in 2000 January. The light curves' properties
were chosen as in George et al. (1998a) to enable comparison to their
Figure~1. The adopted bin size is 5760~s which is one {\it ASCA} orbit
(the good-time interval within one orbit is $\approx 1600$~s). On 2000
January 21 there is a 35~ks gap between the last bin and the preceding
one (see \S~\ref{obsasca} for details). {\em Top and middle:} Summed
light curves obtained for the SIS and GIS detectors. {\em Bottom:}
Softness ratio (the ratio of the summed count rates observed in the
0.5--2~keV and 2--10~keV bands) for the SIS detectors. Variability is
clearly apparent and generally follows the same properties as in George
et al. (1998a).
\label{asca_lc} }
\centerline{}
\centerline{}
%\end{figure}

\noindent In Figure~\ref{asca_lc} we present the 0.5--10~keV SIS and
2--10~keV GIS light curves. Variability is apparent both between and
within individual observations as was found by George et al. (1998a)
for the 1993 and 1996 observations.

During previous {\it ASCA} observations the count rates have varied in
the ranges of 1.4--3.7 ct\,s$^{-1}$ for the SIS and 1.2--3.0 ct\,s$^{-1}$
for the GIS. The count rate variation ranges of the 2000 observations
(2.4--3.7 ct\,s$^{-1}$ for the SIS and 1.3--2.1 ct\,s$^{-1}$ for the
GIS) are within the past ranges. Also the softness ratio of the current
observation, which varies in the range 0.95--1.25, is consistent with
the ranges from the 1993 and 1996 observations (0.8--1.4). The 2000
{\it ASCA} observations suggest that the properties of NGC\,3783 have
remained basically the same, and that during 2000 January
(simultaneously with the {\em Chandra} observation) NGC\,3783 was in a
typical, representative X-ray state.

Light curves for the simultaneous observations of 2000 January 20--21
from the three observatories are shown in Figure~\ref{lcall}. During
the {\it Chandra} observation the AGN shows a rise of $\sim$30\% in the
X-ray flux and then a decrease by about the same amount. The {\it RXTE}
time variations are consistent with the HEG and MEG variations. The
{\it ASCA} variations are also consistent with those from the \ other two \
observatories, \ although the live

%\begin{figure}
\centerline{\includegraphics[width=8.5cm]{f3.eps}}
\figcaption{Light curves from the {\it ASCA} SIS (upward pointing
triangles; 0.5--10 keV), {\it RXTE} PCA (squares; 3--25 keV), {\it
Chandra} MEG (circles; 0.5--7 keV), and {\it Chandra} HEG (stars;
0.5--7 keV). For clarity of presentation the {\it ASCA} SIS, {\it RXTE}
PCA, and {\it Chandra} HEG light curves have been divided by factors of
2, 20, and 0.67, respectively. Also shown is the 0.5--7 keV {\it
Chandra} background light curve (downward pointing triangles) in units
of ct\,s$^{-1}$\,pixel$^{-1}$ which was multiplied by 800000 for
clarity.
\label{lcall} }
\centerline{}
\centerline{}
%\end{figure}

\noindent time of the {\it ASCA} observation is
very short due to the problem noted in \S~\ref{obsasca}. The
zeroth-order HETGS image does not contain useful time variation
information since it is heavily piled-up, although during the rise in
the X-ray flux the pile-up is more severe. For comparison purposes, we
also show in Figure~\ref{lcall} a background light curve extracted from
a region on the ACIS-S S3 CCD. No significant variations are seen in
the background light curve. We have investigated spectral variability
during the {\it Chandra} observation using hardness-ratio light curves
as well as and Kolmogorov-Smirnov tests in wavelength bins with widths
comparable to the spectral resolution. We do not find any strong
spectral variability that might compromise the spectral analysis
below.

\section{Modeling the Continuum}

\subsection{The {\it Chandra} Data}
\label{under_con}

The presence of numerous strong absorption lines, emission lines, and
absorption edges (see below) complicates the determination of the
continuum. To address this issue we introduce the concept of
``line-free zones'' (LFZs) which are spectral regions free of
absorption and emission lines from cosmically abundant elements. The
LFZs are crucial for determining the uncertainty on the underlying
continuum as a function of wavelength, and hence the uncertainty on the
line EWs (\S~\ref{comparison}).

\begin{figure*}[t]
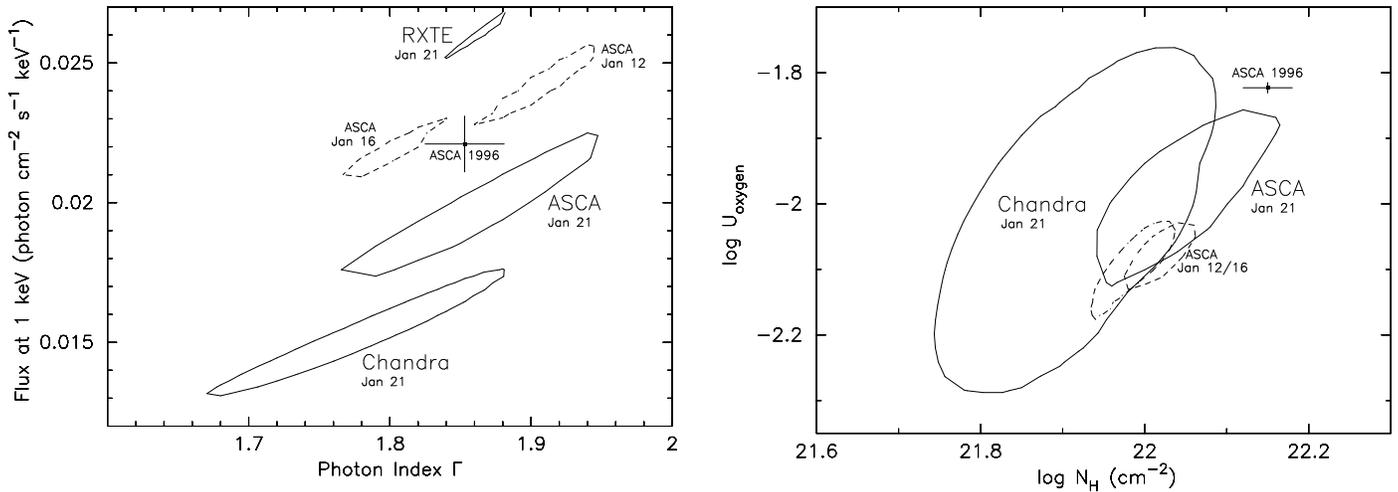

\hglue0.0cm{\includegraphics[angle=-90,width=8.9cm]{f4a.eps}}
\hglue0.5cm{\includegraphics[angle=-90,width=8.9cm]{f4b.eps}}
\caption{The solid curves show the 90\% confidence contours of the
best-fitting parameters associated with the continua as derived using
the {\it Chandra} LFZs (\S\ref{under_con}) and the 2000 January 21 {\it
ASCA} and {\it RXTE} data (\S~\ref{ascaspec}). All three datasets are
consistent with a photon index of $\Gamma \approx 1.85$, an ionized column
density of $N_{\rm H} \approx 10^{22}\ {\rm cm^{-2}}$, and an ionization
parameter of $\log U_{\rm oxygen} \approx {-2}$. The apparent discrepancies
in the absolute fluxes of the different missions are discussed in
\S~\ref{continuum_comp}. Also shown (dashed) are the contours for the two
other {\it ASCA} observations made in 2000, and (crosses) the archival
{\it ASCA} observations from 1996. We stress that the values of
$N_{\rm H}$ and $U_{\rm oxygen}$ shown here are only crude
parameterizations of the true state of the ionized gas (see
\S~\ref{photo_models}).
\label{fig:contours} }
\end{figure*}

We have determined the LFZs for the MEG and HEG data from the line list
used by {\sc ion2000} (the 2000 version of {\sc ion}; see Netzer 1996,
and Netzer, Turner, \& George 1998). These wavelength ranges (in the
rest frame) were then further restricted to allow for intrinsic
velocity shifts of $\pm$600~\kms\ and to account for the resolution of
the MEG (i.e., on each side we reduced the LFZs by a distance, in
wavelength, of 600~\kms\ plus 0.023 \AA ). We also excluded from the
LFZs absorption features which are seen in the data but are not
included in {\sc ion2000} (such as absorption by iron M-shell ions
around 16--17 \AA; see \S~\ref{comparison}). LFZs at energies above
5~keV were excluded to avoid possible contamination by broad Fe
K$\alpha$ emission (\S~\ref{feka}).  The LFZs (green horizontal regions
in Figure~\ref{megspec}) were used on the combined MEG and HEG
1st-order spectrum we derived above in \S~\ref{chandra_obs}. The data
within each LFZ were extracted and rebinned such that each new bin
contained at least 30 counts.  The resulting 315 bins were then fitted
using $\chi^2$-minimization within the {\sc xspec} Version 11.0.1
software (Arnaud 1996). It is important to note that the LFZs provide a
measure of the {\it observed} continuum, which will include spectral
curvature resulting from bound-free absorption edges caused by the warm
absorber and any ``Compton-reflected'' continuum at high energies. The
spectral model adopted for the LFZs therefore consisted of a power-law
continuum (of photon index $\Gamma$), an ionized absorber, and a
reflection continuum from neutral material. For simplicity, and to
enable easy comparison with the {\em ASCA}, {\em RXTE} and previous
work, in this section we {\it parameterize} the ionized absorber using
a single-zone model with a column density $N_{\rm H}$ and an ionization
parameter $U_{\rm oxygen}$ defined over the 0.538--10~keV band (as
described in George et al. 2000). As further discussed in H. Netzer
(in preparation) this is a more suitable quantity for parameterizing the
state of the ionized gas visible in the {\it Chandra} and {\it ASCA}
bandpasses. We shall discuss this parameterization in the context of
our full model in \S~\ref{photo_models}. A reflection continuum is
required by the {\it RXTE} data, but it has a negligible effect on the
analysis of the LFZs described here. It is included here for
completeness, with its intensity restricted to the range determined
from the {\it RXTE} data (see \S\ref{ascaspec}).

We find such a model provides an excellent description of the LFZ data
($\chi^2_{\nu} = 0.81$ for 310 degrees of freedom; dof). The
best-fitting parameters are $\Gamma=1.77^{+0.10}_{-0.11}$, a
normalization of $A$(1~keV) $ = (1.51^{+0.26}_{-0.22})\times 10^{-2}\
{\rm photon\ cm^{-2}\ s^{-1}\ keV^{-1}}$, $\log N_{\rm H} =
21.94^{+0.15}_{-0.20}$ and $\log U_{\rm oxygen} =
-2.01^{+0.25}_{-0.28}$. The 90\% confidence contours of these
parameters are shown in Figure~\ref{fig:contours}, along with those for
the other data sets discussed below. We have determined the uncertainty
on the observed continuum by fixing each parameter in turn at both its
minimum and maximum value consistent with the data at 90\% confidence.
The spectral analysis was then repeated for each limit on each
parameter and the ``extreme'' spectra determined. These extreme spectra
were used to estimate the uncertainties on the line EWs discussed in
\S~\ref{comparison}.

\subsection{The {\it ASCA} and {\it RXTE} Data}
\label{ascaspec}

The {\it ASCA} and {\it RXTE} observations carried out on 2000 January
21 were largely simultaneous with the {\it Chandra} observation (see
Table~\ref{observations} and Figure~\ref{lcall}). For {\it ASCA} all
GIS data below 1~keV were ignored to avoid the uncalibrated degradation
of the detectors late in the mission. The SIS data above 0.6~keV were
included in the analysis, with their poorly calibrated degradation
parameterized using the method outlined in Yaqoob
(2000a).\footnote{Specifically we fixed the ``excess-$N_{\rm H}$'' of
SIS0 at $7.0\times10^{20}\ {\rm cm^{-2}}$, allowing that for SIS1 to be
free during the fitting. The value so derived for the excess-$N_{\rm
H}$ of SIS1 was $(10.7^{+2.6}_{-1.7})\times10^{20}\ {\rm cm^{-2}}$,
consistent with Figure 5 of Yaqoob (2000a).}

For the analysis presented in this section, the data from both
satellites in the 5--7~keV band were again ignored to avoid possible
contamination by broad Fe K$\alpha$ emission (see \S~\ref{feka}). The
spectral model described in \S~\ref{under_con} (also including Galactic
absorption) was applied separately to the {\it ASCA} and {\it RXTE}
data sets. For simplicity, we limit our parameterization of any
reflection component to that from neutral material (Magdziarz \&
Zdziarski 1995) illuminated by an isotropic source above a
disk. The intensity of this component is parameterized by
the inclination angle, $i$, at which the disk is viewed ($i=0\degr$ for a
face-on disk). Acceptable fits for both data sets were obtained.

The bandpass of the {\it RXTE} PCA is 3--25~keV, and hence the PCA data
are unable to constrain effectively either the ionized absorber or any
high-energy cut-off in the underlying continuum. The parameters of the
former were therefore fixed to those determined by {\it ASCA} (see
below), and the cut-off energy was fixed to 200~keV. With these
constraints the best-fit model to the PCA has $\chi^2_{\nu} = 0.93$ (41
dof), $\Gamma=1.86^{+0.03}_{-0.02}$, $A$(1~keV) $ =
(2.60^{+0.10}_{-0.08})\times 10^{-2}\ {\rm
photon\ cm^{-2}\ s^{-1}\ keV^{-1}}$, and $\cos i =
0.46^{+0.22}_{-0.14}$. The {\it ASCA} data are unable to place a useful
constraint on the strength of the reflection component. Thus we
restricted the range of $\cos i$ to be that allowed by the PCA data. We
obtain $\chi^2_{\nu} = 1.02$ (483 dof), $\Gamma=1.88^{+0.10}_{-0.12}$,
$A$(1~keV) $ = (1.99^{+0.32}_{-0.29})\times 10^{-2}\ {\rm
photon\ cm^{-2}\ s^{-1}\ keV^{-1}}$ (for SIS0), $\log N_{\rm H} =
22.06^{+0.11}_{-0.14}$, and $\log U_{\rm oxygen} =
-1.98^{+0.13}_{-0.15}$. These parameters are shown in
Figure~\ref{fig:contours}.

Also shown in Figure~\ref{fig:contours} are the confidence regions
obtained by applying the same spectral model to the other two {\it
ASCA} observations made in 2000 January (dashed curves). Such a model
provided acceptable fits to both of the other 2000 data sets
($\chi^2_{\nu}$/dof $=$ 1.06/1049 and 1.09/1053 for the January 12 and
16 observations, respectively) with best-fitting spectral shape
parameters consistent with those obtained from the January 21 data.
The weighted mean values derived from these three {\it ASCA}
observations are $\Gamma=1.86\pm0.07$, $\log N_{\rm H} = 22.01\pm0.03$,
and $\log U_{\rm oxygen} = -2.07\pm0.04$.

We have applied the same spectral model to all archival {\it ASCA}
observations of NGC~3783. This analysis differs from that presented in
George et al. (1998b) due to (1) the explicit inclusion of the
reflection component, (2) the use of the Yaqoob (2000a) method to
account for the degradation of the SIS detectors, (3) the assumption of
a single power-law photoionization continuum over the entire
0.1--200~keV band, and (4) the use of $U_{\rm oxygen}$ (0.538--10~keV)
rather than $U_{\rm x}$ (0.1--10~keV). Taking into account the turn-up
in the spectrum below 0.2~keV assumed by George et al. (1998b), $U_{\rm
oxygen} = 0.114\,U_{\rm x}$ for $\Gamma = 1.77$ (derived from the {\it
Chandra} spectrum in \S~\ref{under_con}). Our results are generally
consistent with those in George et al. (1998b). We find all four 1996
observations to be consistent with $\Gamma=1.85\pm0.03$, $A$(1~keV) $ =
(2.21\pm0.10)\times 10^{-2}\ {\rm photon\ cm^{-2}\ s^{-1}\ keV^{-1}}$,
$\log N_{\rm H} = 22.15\pm0.03$, and $\log U_{\rm oxygen} =
-1.82\pm0.01$. These are shown as crosses in Figure~\ref{fig:contours}.
The two {\it ASCA} observations performed in 1993 have
$\Gamma=1.86\pm0.01$ and $A$(1~keV)$ = (1.6$--$1.9)\times 10^{-2}\ {\rm
photon\ cm^{-2}\ s^{-1}\ keV^{-1}}$. The \ion{O}{7} and \ion{O}{8}
edges are deeper than those seen at later epochs, and the ionized gas
may be parameterized by $\log N_{\rm H} \approx 22.3$ and $-1.9
\lesssim \log U_{\rm oxygen} \lesssim -1.8$. However, as found by
George et al. (1998b), the spectra at this epoch cannot be adequately
modeled by a single-zone ionized absorber.

\subsection{Continuum Comparison Between the Satellites}
\label{continuum_comp}

The best-fitting values for $\Gamma$ from the {\it ASCA} and {\it RXTE}
data sets ($\Gamma = 1.88^{+0.10}_{-0.12}$ and
$\Gamma=1.86^{+0.03}_{-0.02}$, respectively, obtained in
\S~\ref{ascaspec}) are consistent, and they are also consistent with
that obtained for the {\it Chandra} LFZs in \S~\ref{under_con}
($\Gamma=1.77^{+0.10}_{-0.11}$). The best-fitting values of $N_{\rm H}$
and $U_{\rm oxygen}$ using the {\it Chandra} LFZs and the {\it ASCA}
data are also consistent (see Figure~\ref{fig:contours}). We find
significant differences, however, in the values derived for $A$(1~keV).
Presumably these partly result from differences in the absolute flux
calibrations of the three missions. Specifically we find the values of
$A$(1~keV) derived from the {\it ASCA} SIS0, SIS1 and GIS2 detectors to
agree within 1\%, while that for GIS2 is $\sim 8$\% lower. These values
are therefore consistent with the quoted uncertainties on the mean
absolute flux calibration ($\lesssim 3$\%) and internal
detector-detector fluctuations ($\lesssim 10$\%; Yaqoob 2000b). The
value of $A$(1~keV) derived using the {\it RXTE} PCA data is $\sim$30\%
higher than for the SIS0. Such a value is consistent with that expected
from the known offset in the absolute calibrations as determined from
simultaneous observations of 3C~273; for these the PCA was 30--40\%
higher than {\it ASCA} and {\it BeppoSAX} (Yaqoob \& Serlemitsos 2000;
Yaqoob 2000b).

The value of $A$(1~keV) derived using the LFZs and the {\it Chandra}
HETGS data is $\approx$25\% lower than for the SIS0. To compare better
between the {\it Chandra} and {\it ASCA} data we binned the {\it
Chandra} spectrum to the {\it ASCA} resolution and overplotted it with
the {\it ASCA} model derived for the 21 January 2000 observation (upper
panel in Figure~\ref{comparison_e}). The two data sets have good
overall agreement in shape, and the systematic shift between them is
clearly seen. \ The average ratio \ between the \ {\it Chandra}  

%\begin{figure}
\centerline{\includegraphics[width=8.5cm]{f5.eps}}
\figcaption{{\it Upper panel:} A comparison of the {\it Chandra} spectrum
and the {\it ASCA} model obtained for the simultaneous observation of
21 January 2000. The {\it Chandra} data are binned to the {\it ASCA}
resolution and are shown as dots with error bars where the horizontal
error bar denotes the bin size and the vertical error bar denotes the
rms of the data averaged in the bin.  Note that the {\it Chandra} data
are consistent with the presence of ionized edges from \ion{O}{7} and
\ion{O}{8}.  {\it Lower panel:} The ratio between the {\it Chandra}
data and the {\it ASCA} model presented in the upper panel. The
systematic $\approx$~15\% difference is noticeable as well as the 1~keV
deficit feature (see \S~\ref{continuum_comp} for details).
\label{comparison_e}}
\centerline{}
\centerline{}
%\end{figure}

\noindent data and the
{\it ASCA} model is $0.87\pm0.01$ over the 0.5--7 keV range (lower
panel in Figure~\ref{comparison_e}; the quoted error is the standard
deviation of the mean). The discrepancy is somewhat stronger in the
0.5--2 keV band (ratio of $0.82\pm 0.02$) than in the 2--7 keV band
(ratio of $0.90\pm 0.01$). These ratio values are consistent with the
{\it Chandra} flux calibration uncertainties in the two bands (see
\S~\ref{chandra_obs}). However, the shape of the {\it Chandra} data to
{\it ASCA} model ratio along the energy axis suggests this difference
between the hard and soft bands might also be a manifestation of the 1
keV deficit feature noted by George et al.  (1998a); we further discuss
this feature in \S~\ref{comparison}. The average ratio of the {\it
Chandra} data to {\it ASCA} model is smaller than the $A$(1~keV)
difference of $\approx$25\% stated above. We suggest that the somewhat
different values of $N_{\rm H}$, $U_{\rm oxygen}$, and $\Gamma$ between
the two data sets contribute to lower the discrepancy seen in
$A$(1~keV) by about a factor of 2. This is also seen when comparing the
total fluxes measured from the two data sets (after correcting for
Galactic absorption). The {\it Chandra} observation yields $(2.0\pm0.2)
\times 10^{-11}$ \ergcms\ in the 0.5--2~keV rest-frame band and
$(4.1\pm 0.1) \times 10^{-11}$ \ergcms\ in the 2--7 keV rest-frame
band, while the {\it ASCA} observation yields $(2.3\pm 0.2 ) \times
10^{-11}$ \ergcms\ and $(4.5^{+0.1}_{-0.2}) \times 10^{-11}$
\ergcms\ in the two bands, respectively (the uncertainties were
computed using the extreme spectra described at the end of
\S~\ref{under_con}). The differences in the fluxes between the two
instruments is $\la 15\%$.

It should be noted that the source exhibits intensity variations during
the observations ($\pm 15$\%; Figure~\ref{lcall}). Since the
time-averaged {\it ASCA}, {\it Chandra} and {\it RXTE} spectra sample
slightly different times, this may account for at least some of the
discrepancies in $A$(1~keV) (e.g., the mean count rate over the whole
{\it RXTE} observation is $\sim6$\% higher than that during the first
two {\it ASCA} orbits). We stress that these uncertainties in the flux
calibration do not affect the main conclusions of this paper. 

The best-fitting continuum models to both the {\it Chandra} and {\it
ASCA} data clearly require \ion{O}{7} and \ion{O}{8} absorption edges.
While these edges, previously found in {\it ASCA} data of NGC\,3783 and
many other AGNs, are not clear on the scale of Figure~\ref{megspec},
their influence on the power-law continuum is clearly present in the
data, especially in view of the good agreement between the shape of the
{\it ASCA} model and the {\it Chandra} data (see the top panel of
Figure~\ref{comparison_e}). However, as seen in Figure~\ref{megspec}, the
high-resolution X-ray spectrum of NGC\,3783 has many more features
which are blended with these edges.

\section{Analysis and Modeling of the {\it Chandra} Data}
\label{spectral}

We have modeled the high-resolution {\it Chandra} spectra using the
basic technique described in Paper~I but focusing, this time, on (1)
individual spectral features, (2) the line profiles, shifts and EWs,
(3) the general physical model, and (4) the spectral energy
distribution (SED).  The following is a detailed account of the major
findings starting with the empirical determination of the
absorption-line widths and shifts and ending with global emission and
absorption models.

\subsection{New Measurements of Absorption Lines}

\subsubsection{Direct Constraints on the Absorption-Line Widths}
\label{absorption_lines}

There are two ways to obtain the absorption-line widths and their
velocity field: (1) direct measurement from the data and (2) a
comparison of the observed and modeled EWs (deducing the velocity from
the model). Paper~I used the second method; this is re-evaluated in
\S~\ref{comparison}. Here we demonstrate that the high-resolution
spectrum is of high enough quality to resolve some of the absorption
lines and obtain direct line-width measurements.

The spectral resolutions of the {\it Chandra} gratings are
approximately constant with wavelength: 0.023~\AA\ for the MEG and
0.012~\AA\ for the HEG (FWHM of Gaussian profiles). At 15 \AA, these
correspond to $\approx 460$~km\,s$^{-1}$ for the MEG and $\approx
240$~km\,s$^{-1}$ for the HEG. In Paper~I we were not able to constrain
directly the velocity dispersion of the absorbing gas because of the
poor S/N of individual absorption lines. We have therefore tried a
different method of adding several absorption lines from the same
element. The advantage is the significant increase in S/N. The
disadvantage, the somewhat deteriorated spectral resolution since, for
each element, the resolution is then defined by the line with the
shortest wavelength.

\begin{figure*}[t]
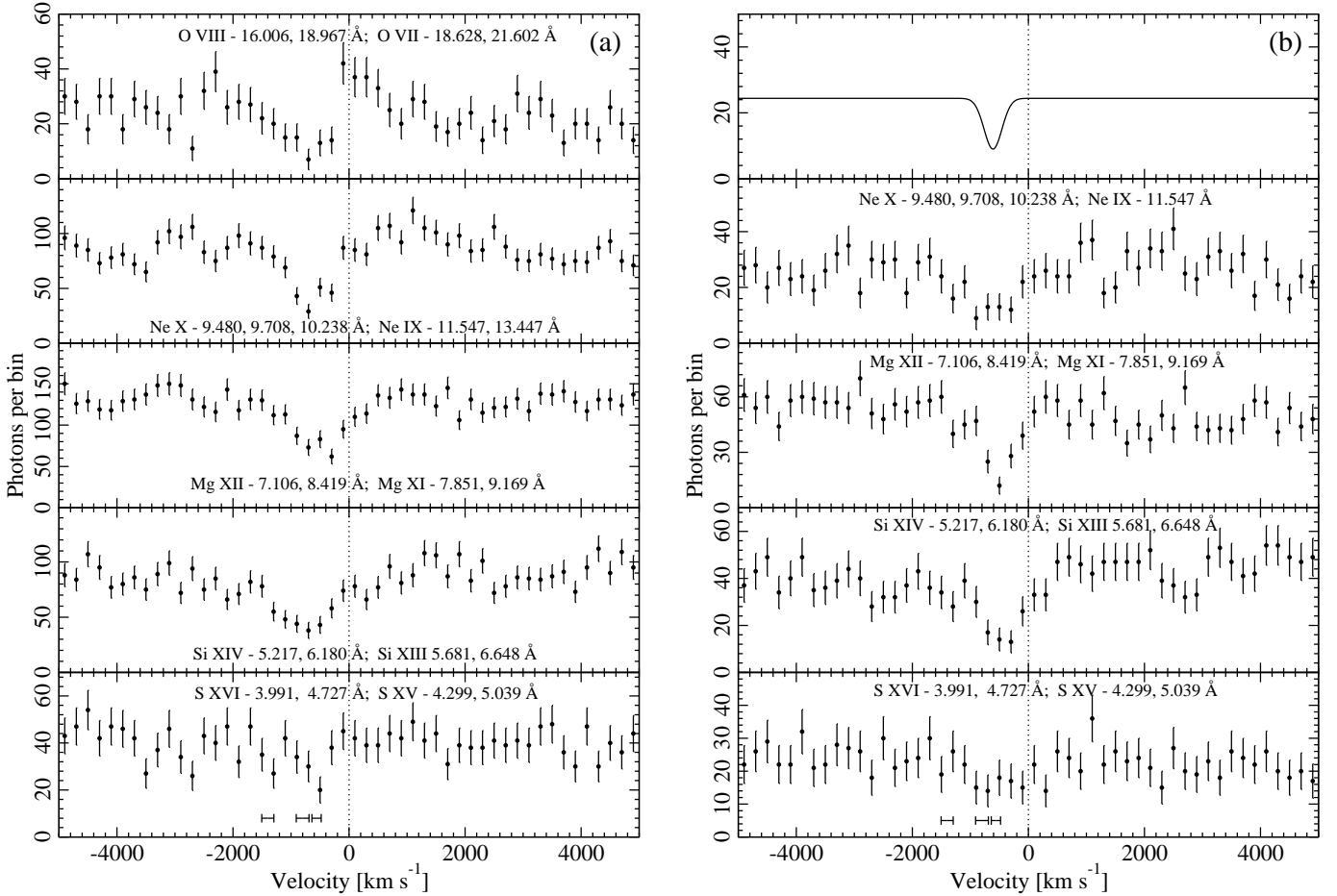

\hglue0.0cm{\includegraphics[width=8.8cm]{f6a.eps}}
\hglue0.5cm{\includegraphics[width=8.8cm]{f6b.eps}}
\caption{(a) MEG and (b) HEG velocity spectra showing co-added lines
from oxygen, neon, magnesium, silicon and sulfur. The bin size is 200
km\,s$^{-1}$. The wavelengths of the lines co-added for each element
are listed in each panel. Error bars have been computed following
Gehrels (1986). Note that absorption is clearly detected from oxygen,
neon, magnesium and silicon. There is also a hint of absorption by
sulfur, although this is not significant.  Note also the P~Cygni
profile for the oxygen lines.  For the HEG we have not shown the
co-added velocity spectrum for oxygen, as the photon statistics are
poor. Instead, we show a Gaussian absorption line representing the line
response function of the HEG at 9.480\,\AA\ (the Gaussian FWHM is
364\,km\,s$^{-1}$); this is the poorest line response function
applicable to the co-added velocity spectrum for neon, shown in the
panel immediately below. Note that the neon absorption line appears
significantly broader than the line response function.
The velocity shifts and dispersions of the UV absorption systems are
marked as three horizontal lines in the lower panels.
\label{vel_spec} }
\end{figure*}

We created ``velocity spectra'' by adding, in velocity space, several
absorption lines from the same element. The lines were chosen to be the
strongest predicted features from a given element and to be free from
contamination by adjacent features. The velocity spectra were built up
on a photon-by-photon basis (rather than by interpolating spectra
already binned in wavelength). The velocity spectra shown in
Figure~\ref{vel_spec} demonstrate the clear detection of absorption
lines from oxygen, neon, magnesium and silicon
(the ions and lines which are used to create the spectra are
listed in the figure).
The detection of sulfur absorption is marginal.

We fitted each of the velocity spectra with a Gaussian profile whose
centroid and width are free parameters. The resulting centroids and
widths ($\sigma$) are listed in Table~\ref{linewidth} columns (3) and
(4) (the insignificant fit of the sulfur line was omitted from the
table).  Column (5) of Table~\ref{linewidth} lists the expected
instrumental $\sigma$ ($\sigma_{\rm instrument}$) for each element
computed using the instrumental FWHM resolution (here and elsewhere
FWHM=$2.3548\sigma$). The list conservatively refers to the most poorly
resolved line, i.e., the one with the shortest wavelength (see
Figure~\ref{vel_spec} for all wavelengths used). For the HEG profiles,
$\sigma_{\rm instrument}$ is consistent with $\sigma_{\rm measured}$
for all lines except the neon lines that are apparently resolved.
Assuming $\sigma_{\rm true}^2=\sigma_{\rm measured}^2-\sigma_{\rm
instrument}^2$, we have proceeded to obtain $\sigma_{\rm true}$, or its
upper limit. This is listed in column (6) of Table~\ref{linewidth}. A
few values of $\sigma_{\rm true}$ have upper limits which are lower
than the value we find for the neon lines. This may be a result of the
conservative instrumental FWHM resolution we used. However, {\it all}
values of $\sigma_{\rm true}$ are consistent with $\approx 210$~\kms\
(this value will be used for our model in \S~\ref{photo_models}).  For
the neon lines we find from the HEG $\sigma_{\rm true} =
360^{+210}_{-150}$ km\,s$^{-1}$ (i.e., FWHM$_{\rm
true}=840^{+490}_{-360}$ km\,s$^{-1}$). Comparing the $\chi^2$ values
of our fits to the HEG neon lines we find a $\Delta\chi^2$ of 9.3 when
allowing the line to be broad (rather than fixing it at the instrumental
resolution). According to the $F$-test for 95 dof, this $\Delta\chi^2$
corresponds to a highly significant improvement in fit quality at
$>99.5$\% confidence (see Table~C.5 of Bevington \& Robinson 1992).

To check for possible problems, we have created two independent
velocity spectra for neon from the HEG~$+1$st and $-1$st orders. Both
are consistent with the presence of line broadening and are consistent
with each other. For the HEG~$+1$st order, $\sigma_{\rm measured}=
374^{+408}_{-193}$ km\,s$^{-1}$ (corresponding to $\sigma_{\rm
true}=340^{+427}_{-245}$ km\,s$^{-1}$), and for the HEG~$-1$st order,
$\sigma_{\rm measured}= 390^{+313}_{-186}$ km\,s$^{-1}$ (corresponding
to $\sigma_{\rm true}=359^{+327}_{-224}$ km\,s$^{-1}$). As an
additional test, we measured the width of the readout trace from the
zeroth order image. The PSF of the readout trace (which consists of
$\sim 550$ counts) has a FWHM of $2.07\pm 0.20$ pixels. This is broader
by 0.5--0.7 pixels than the FWHM of 1.33 pixels reported in the {\it
Chandra Proposers' Observatory Guide} or the $1.554\pm 0.013$ pixels
derived from a sample of point sources (Sako et al. 2000). We are
unable to fully explain the origin of this small discrepancy. It may
result from small inaccuracies in the {\sc dmregrid} software used for
the image processing. If this 0.5--0.7 pixel broadening indicates a
problem in the aspect reconstruction, it will amount to a velocity
broadening of $\sigma$ = 88--123~km\,s$^{-1}$ given the HEG resolution
at 9.480 \AA\ (the most poorly resolved neon line we consider). This is
small compared with the derived $\sigma$ of neon and does not change
the conclusion that the neon lines are resolved. Accounting for such a
broadening we obtain FWHM$_{\rm true} =
(792^{+511}_{-402})$--$(817^{+501}_{-378})$ km\,s$^{-1}$.

All line centroids listed in Table~\ref{linewidth}, column (3), are
clearly blueshifted with respect to the systemic velocity. The mean and
rms blueshift are $-610\pm 130$ km\,s$^{-1}$. This is consistent with
the blueshifts of all the individual features in Table~\ref{lines} (see
below).

\subsubsection{X-ray P~Cygni Profiles}

The velocity spectrum of oxygen (the top panel in
Figure~\ref{vel_spec}a) also reveals the existence of an emission
feature. A half-Gaussian fit to this feature yields a peak at
$-100\pm200$ km\,s$^{-1}$ and a Gaussian $\sigma_{\rm measured} =
486^{+237}_{-159}$ km\,s$^{-1}$ (FWHM$_{\rm true} = 1060^{+585}_{-422}$
km\,s$^{-1}$). Thus, we find indications of a P~Cygni type profile with
an overall width consistent with twice the mean blueshift of the
absorption lines. This is the first clear detection of an X-ray P~Cygni
profile from an extragalactic source and one of the first P~Cygni
profiles in X-ray astronomy (see Brandt \& Schulz 2000). This profile
will be used as input to our global emission and absorption model.

%\begin{deluxetable}{cccccc}
%\tablecolumns{6}
%\tablewidth{0pt}
\begin{table*}
\footnotesize    % better fit the Journal font size.
\caption{Fitting of Added Absorption-Line Profiles\tablenotemark{a}
\label{linewidth}}
\begin{center}
\begin{tabular}{cccccc}
\hline
\hline
{} &
{} &
{Line centroid} &
{$\sigma_{\rm measured}$} &
{$\sigma_{\rm instrument}$} &
{$\sigma_{\rm true}$} \\
{Element} &
{Spectrum} &
{(km\,s$^{-1}$)} &
{(km\,s$^{-1}$)} &
{(km\,s$^{-1}$)} &
{(km\,s$^{-1}$)} \\
{(1)} &
{(2)} &
{(3)} &
{(4)} &
{(5)} &
{(6)} \\
\hline
O  & MEG &$-715^{+132}_{-151}$ &$282^{+131}_{-105}$& 183 & $<370$ \\ [0.02cm]
Ne & HEG &$-685^{+178}_{-181}$&$390^{+197}_{-132}$& 155 & $358^{+209}_{-152}$ \\ [0.02cm] 
Ne & MEG &$-668^{+68\phn}_{-69\phn}$ &$292^{+91\phn}_{-77\phn}$& 309 & $<226$ \\ [0.02cm] 
Mg & HEG &$-502^{+64\phn}_{-63\phn}$ &$217^{+81\phn}_{-56\phn}$& 206 & $<215$ \\ [0.02cm] 
Mg & MEG &$-493^{+89\phn}_{-91\phn}$ &$420^{+99\phn}_{-81\phn}$& 412 & $<315$ \\ [0.02cm] 
Si & HEG &$-461^{+103}_{-121}$ &$376^{+163}_{-113}$& 281 & $<459$ \\  [0.02cm]
Si & MEG &$-777^{+116}_{-123}$ &$562^{+182}_{-145}$& 562 & $<488$ \\ 
\hline
\end{tabular}
\vskip 2pt
\parbox{3.5in}{ % use this to define the width of the notes under the table
\small\baselineskip 9pt
\footnotesize
\indent
$\rm ^a${Corresponds to Figure~\ref{vel_spec}.} \\
Note. --- {Uncertainties are 90\% confidence for one parameter
of interest ($\Delta\chi^2~=~2.71$).}
}
\end{center}
\end{table*}
\normalsize

We note that the averaging procedure above, consisting of several lines
from each series, tends to reduce the contrast between the emission and
absorption features for all lines with large optical depths; the
absorption EW is roughly constant for all such lines while the emission
EW is always the largest for the first line in the series (e.g., the
Ly$\alpha$ line in the H-like ions). This is the reason why only the
strongest emission lines, those due to oxygen, are seen in the velocity
spectra. However, we detect individual P~Cygni line profiles in other
elements [e.g., \ion{Ne}{10} (12.132 \AA), \ion{Si}{13} (6.648 \AA); see
Figure~\ref{megspec}].

%\begin{figure}
\centerline{\includegraphics[width=8.7cm]{f7.eps}}
\figcaption{ The broken power-law continuum used in the modeling described
in \S~\ref{photo_models} and detailed in Table~4. The {\it
Chandra} and {\it HST STIS} observable ranges are marked with
double-headed arrows. The unobservable part of the continuum due to the
Galactic column density is between the two vertical dashed lines;
see~\S~\ref{uv_absorber}.
\label{SED} }
%\end{figure}

\subsection{Photoionization Models}
\label{photo_models}

\subsubsection{Model Parameters and Line Widths}

We proceed by calculating a number of photoionization models that are
compared, in turn, to the observed spectrum. The basic assumptions are
outlined in Paper~I, but we list them here for completeness.

All models consider one or more outflowing shells of constant density,
solar metallicity, highly ionized gas (HIG). We assume this gas is in
photoionization and thermal equilibrium. The HIG is illuminated by a
central broken power-law continuum of photon index $\Gamma$ as defined
in Table~4 and illustrated in Figure~\ref{SED}. The choice
of the 0.1--50 keV slope of $\Gamma=1.77$ is the result of
experimenting with a range of models, as explained in
\S~\ref{under_con} and later in this section. The very sharp decline
over the 40--100 eV range is consistent with the expected $\sim 10^5$~K
accretion disk temperature. The other parameters of the model are the
column density, covering fraction and ionization parameter for each
shell.

In Paper~I we used the 0.1--10 keV ionization parameter $U_{\rm x}$.
For a simple power-law spectrum with $\Gamma=1.77$, $U_{\rm oxygen}=
0.252\,U_{\rm x}$. We also define the internal microturbulence velocity
which is added, in quadrature, to the locally computed thermal velocity
of the ions, $V^2=V^2_{\rm thermal}+V^2_{\rm turbulence}$ (these are
Doppler profiles; i.e., the velocity $V$ is given by
$V=\sqrt{2}\sigma$). This defines the line widths and is an important
factor which determines the line EWs. We note that $V_{\rm thermal} \ll
V_{\rm turbulence}$ in our model as the temperature of the gas is
$\approx 10^5$ K.

The model of Paper~I included a single-component HIG with a large
covering fraction, a column density of $\log N_{\rm H}=22.1$, solar
metallicity, and an ionization parameter of $U_{\rm x}=0.13$ ($U_{\rm
oxygen}=0.01$). A closer examination of this model shows that, while
the fit to the continuum level and the \ion{O}{7} and \ion{O}{8}
bound-free absorption features is satisfactory, there are three
fundamental problems: (1) the assumed microturbulence velocity of
150~\kms\ underpredicts the EWs of several neon and magnesium
absorption lines, (2) the EWs of several high-energy absorption lines
[e.g., \ion{Si}{14}~(6.180~\AA)] are underpredicted by factors of at
least 3, and (3) the model made no attempt to explain the rich spectrum
of iron L-shell lines.

Regarding point (1), we note that the calculations presented in Paper~I
contained a factor of 1.7 error in the model EW calculations. Thus, the
EWs quoted there as corresponding to a microturbulence velocity of
150~\kms\ correspond, in fact, to a microturbulence velocity of about
250 \kms. The more detailed fitting in \S~\ref{absorption_lines}
indicates that even a somewhat larger velocity is likely. In the
following we adopt as our standard $V_{\rm turbulence}=300~\kms$
($\sigma \approx 210~\kms$). Regarding point (2), several attempts to
change the microturbulence velocity, the column density, and the
ionization parameter show that no model can explain the EWs of the
high-ionization species and, at the same time, keep the good fit of the
EWs of many lower ionization lines. We are driven to the conclusion
that modeling the observed spectrum requires more than one absorbing
component. In \S~\ref{two_component} we describe our attempts to fit a
multi-component model to the spectrum of NGC~3783. As for point (3), we
have considerably modified {\sc ion} (into {\sc ion2000}) to include
hundreds of new iron lines, as explained in \S~\ref{Fe_L}.

\subsubsection{The Iron L-Shell Lines}
\label{Fe_L}
\vglue -0.1cm

{\sc ion2000} includes substantial improvements to its atomic data base
designed to model the numerous iron L-shell absorption features
apparent in the spectrum. Atomic calculations have been performed using
the multi-configuration, relativistic Hebrew University Lawrence
Livermore Atomic Code ({\sc hullac}) developed by Bar-Shalom, Klapisch,
\& Oreg (2001). In this code, the energy levels are calculated with the
relativistic version of the parametric potential method (Klapisch et
al. 1977), including configuration mixing. Subsequently, the oscillator
strengths for the radiative transitions are computed using first-order
perturbation theory.

Photon impact processes involving L-shell (i.e., the $n=2$ electronic
shell, where $n$ is the principal quantum number of the electrons in
the outermost occupied orbital of the ion in its ground state)
excitations are computed for \ion{Fe}{17} through \ion{Fe}{24}. All of
the important transitions from $n$ to $n^\prime$
($n^\prime$\,{$\geq$}\,3) are taken into account, of which the 2p--3d
excitations are the strongest and, therefore, those clearly observed in
the spectrum. Only excitations from the ground level of each shell are
included, except for \ion{Fe}{19}, in which the first excited level can
be significantly populated even at relatively low densities.
Consequently, absorption processes from that excited level of
\ion{Fe}{19} are also calculated. This feature of \ion{Fe}{19} can
potentially provide a density diagnostic at densities of order $10^9$
\cc\ (the precise density is somewhat uncertain due to uncertain atomic
data), since the EWs for absorption lines from this excited level
depend strongly on the density of the absorbing medium (see
\S~\ref{comparison}).

The new set of oscillator strengths obtained in this way are used to
compute the optical depths of several hundred iron L-shell lines
contained in {\sc ion2000}. In two recent works, similar atomic data
for the Fe L-shell transitions were used for line-by-line analysis of
high-resolution spectra of emission (Behar, Cottam, \& Kahn 2001) and
absorption (Sako et al. 2001) lines. In those studies good agreement
was found between the calculations and the observations.

\vglue -0.2cm
\subsubsection{Two-Component Models}
\label{two_component}
\vglue -0.1cm

We have experimented with several two-component models. These include
two distinct shells situated at different radial locations that are
characterized by different column densities and ionization parameters.
Guided by the good continuum fit (\S~\ref{under_con}), we have explored
the following range of parameters (which is not necessarily unique):
$\Gamma=1.77 \pm 0.1$, $\log N_{\rm H}={22.1\pm 0.2}$, and $\log U_{\rm
oxygen}$ from $-1.824$ to $-0.700$. A possible combination that gives
good agreement with the observations (see \S~\ref{comparison}) is made
of the following components:
\begin{description}
\item[The low-ionization component] with $\log U_{\rm oxygen}=-1.745$, a
hydrogen column density of $\log N_{\rm H}=22.2$, a global covering
factor of 0.5 and 
a line-of-sight covering factor of unity.
\item[The high-ionization component] with $\log U_{\rm oxygen}=-0.745$, a
hydrogen column density of $\log N_{\rm H}=22.2$, a global covering factor of
0.3, and a line-of-sight covering factor of unity.
\end{description}
The gas density in both components is assumed to be $10^{8}$~\cc. This
density is high enough so that $\Delta R \ll R$, where $R$ is the
distance to the continuum source and $\Delta R$ is the shell thickness.
We have no real constraints on the density (except that it must exceed
about $ 10^4$~\cc\ to justify the thin shell approximation) and hence
no way to calculate the masses of these components. Under our
assumptions the low-ionization component is further away from the
source and is thus illuminated by the radiation passing through the
high-ionization shell.  However, the high-ionization component is
almost transparent to continuum radiation from the source, and due to
the spectrally distinct line features of the two components, we can
neglect the changes in the illuminating spectrum for the (outer)
low-ionization shell. The two components are assumed to have the same
outflow and microturbulence velocities which is consistent with the
observations. The EW calculations listed in \S~\ref{comparison} are
made under the assumption of no velocity shift between the two
components.

\subsubsection{Continuous Flow Models}
\label{con_flow}

We have also tested continuous models made out of a single outflowing
component of HIG. In this case, geometrical dilution is important, and
different radial locations are characterized by different ionization
parameters. We assumed a density law of the form $N(R) \propto
R^{-\alpha}$ with $\alpha < 2$. Thus the ionization level decreases
outward, and the global properties resemble the two distinct component
model. The requirement of achieving a factor $\ga 10$ reduction in
$U_{\rm oxygen}$, due to geometrical dilution, with a restricted range
in column density, introduces a constraint on the value of $N_0 R_0$
where the subscript zero refers to conditions at the base of the flow.
This results in a low density gas whose total mass exceeds, by several
orders of magnitude, the masses of the two distinct high density shells
in \S~\ref{two_component} (the mass is roughly proportional to $1/N$).

While there are several combinations of $\alpha$ and minimum distance
that satisfy these conditions and give the required range of
ionizations, the overall fit to the data for the combinations we have tried is
inferior to the fit of the two-component model. We have therefore
decided not to present these fits in detail. We emphasize that models
of this kind need further investigation, but this is beyond the scope of
this paper.

%\begin{deluxetable}{cccrl}
%\tablecolumns{5}
%\tabletypesize{\scriptsize}
%\tablewidth{0pt}
\begin{table*}
\footnotesize    % better fit the Journal font size.
\caption{Absorption and Emission Lines from NGC\,3783\tablenotemark{a}
\label{lines}}
\vglue -2cm
\begin{center}
\begin{tabular}{cccrl}
\hline
\hline
{Measured $\lambda$\tablenotemark{b}} &
{Model $\lambda$} &
{Measured EW} &
{Model EW} &
{Ion name and} \\
{(\AA)} &
{(\AA)} &
{(m\AA)} &
{(m\AA)} &
{transition rest-frame wavelength (\AA )} \\
{(1)} &
{(2)} &
{(3)} &
{(4)} &
{(5)} \\
\hline
 \phn4.717 &  \phn4.718 & $    7.8\pm    9.3$ & $   4.5$ &     \ion{S}{16} (4.727)                                                             \\ [-0.013cm]
 \phn5.029 &  \phn5.028 & $\phn8.2\pm   11.3$ & $   8.2$ &     \ion{S}{15} (5.039)                                                             \\ [-0.013cm]
 \phn6.171 &  \phn6.170 & $   17.9\pm    5.1\phn$ & $  11.2$ &     \ion{Si}{14} (6.180)                                                            \\ [-0.013cm]
 \phn6.636 &  \phn6.637 & $   11.1\pm    4.5\phn$ & $  16.5$ &     \ion{Si}{13} (6.648)                                                            \\ [-0.013cm]
 \phn6.681 &  \phn6.668 & $  -10.6\pm    9.1\phm{-1}$ & $  -1.5$ &     \ion{Si}{13} (6.648)                                                            \\ [-0.013cm]
 \phn6.707 &  \phn6.711 & $    5.9\pm    5.0$ & $  11.4$ &     \ion{Si}{12} (6.720)                                                            \\ [-0.013cm]
 \phn7.094 &  \phn7.094 & $    2.9\pm    5.7$ & $   6.0$ &     \ion{Mg}{12} (7.106)                                                            \\ [-0.013cm]
 \phn7.163 &\nodata & $    6.5\pm    2.6$ & \nodata  &     \ion{Al}{13} (7.18)\tablenotemark{c}                                             \\ [-0.013cm]
 \phn7.462 &  \phn7.461 & $\phn4.5\pm   10.1$ & $   9.8$ &     \ion{Mg}{11} (7.473)                                                            \\ [-0.013cm]
 \phn7.744 &\nodata & $    3.8\pm    4.3$ & \nodata  &     \ion{Al}{12} (7.76)\tablenotemark{c}                                             \\ [-0.013cm]
 \phn7.838 &  \phn7.837 & $    7.5\pm    4.7$ & $  15.2$ &     \ion{Mg}{11} (7.851)                                                            \\ [-0.013cm]
 \phn8.403 &  \phn8.404 & $   24.3\pm    5.6\phn$ & $  15.6$ &     \ion{Mg}{12} (8.419)                                                            \\ [-0.013cm]
 \phn9.156 &  \phn9.153 & $   26.5\pm    9.2\phn$ & $  26.4$ &     \ion{Mg}{11} (9.169)                                                            \\ [-0.013cm]
 \phn9.305 &  \phn9.293 & $ -9.1\pm   14.3    $ & $  -6.9$ &     \ion{Mg}{11} (9.300)                                                            \\ [-0.013cm]
 \phn9.356 &  \phn9.327 & $    9.5\pm    9.0$ & $  10.8$ &     \ion{Ne}{10} (9.362), \ion{Mg}{10} (9.336)                                     \\ [-0.013cm]
 \phn9.460 &  \phn9.466 & $   12.0\pm    6.1\phn$ & $  14.0$ &     \ion{Ne}{10} (9.480), \ion{Fe}{21} (9.483)                                     \\ [-0.013cm]
 \phn9.517 &  \phn9.500 & $  -18.6\pm   17.0\phm{-}$ & $  -2.3$ &     \ion{Ne}{10} (9.480), \ion{Fe}{21} (9.483)                                     \\ [-0.013cm]
 \phn9.692 &  \phn9.692 & $   22.2\pm    8.0\phn$ & $  12.4$ &     \ion{Ne}{10} (9.708)                                                            \\ [-0.013cm]
 \phn9.734 &  \phn9.740 & $   -6.4\pm   12.1$ & $  -3.5$ &     \ion{Ne}{10} (9.708)                                                            \\ [-0.013cm]
 \phn9.984 &  \phn9.985 & $   14.9\pm   16.2$ & $   9.8$ &     \ion{Fe}{20} (9.999, 10.001, 10.006)                                            \\ [-0.013cm]
10.032 & 10.042 & $   17.3\pm   13.7$ & $   6.0$ &     \ion{Fe}{20} (10.040, 10.042, 10.054, 10.060)                                     \\ [-0.013cm]
10.100 & 10.100 & $   14.7\pm   12.0$ & $  10.8$ &     \ion{Fe}{17} (10.112)                                                            \\ [-0.013cm]
10.223 & 10.223 & $   18.2\pm    8.5\phn$ & $  19.6$ &     \ion{Ne}{10} (10.238)                                                            \\ [-0.013cm]
10.275 & 10.264 & $  -23.7\pm   22.1\phn$ & $  -2.4$ &     \ion{Ne}{10} (10.238)                                                            \\ [-0.013cm]
10.345 & 10.317 & $   12.8\pm   13.0$ & $   9.7$ &     \ion{Fe}{18} (10.361, 10.363, 10.365)                                            \\ [-0.013cm]
10.500 & 10.488 & $   13.1\pm   12.5$ & $  10.4$ &     \ion{Fe}{17} (10.504)                                                            \\ [-0.013cm]
10.624 & 10.629 & $   46.6\pm   10.9$ & $  20.3$ &     \ion{Fe}{17} (10.657), \ion{Fe}{19} (10.650, 10.642, 10.630, 10.631, 10.641)     \\ [-0.013cm]
10.754 & 10.753 & $   15.9\pm   23.5$ & $  13.7$ &     \ion{Fe}{17} (10.770)                                                            \\ [-0.013cm]
10.798 & 10.809 & $   15.5\pm    8.2\phn$ & $   4.1$ &     \ion{Fe}{19} (10.828)                                                            \\ [-0.013cm]
10.978 & 10.981 & $   43.7\pm   11.7$ & $  53.5$ &     \ion{Ne}{9} (11.000), \ion{Fe}{23} (10.981, 11.019), \ion{Fe}{17} (11.026)       \\ [-0.013cm]
11.119 & 11.116 & $   11.2\pm   20.2$ & $  15.7$ &     \ion{Fe}{17} (1113.9)                                                            \\ [-0.013cm]
11.236 & 11.237 & $   15.7\pm   13.3$ & $  22.0$ &     \ion{Fe}{17} (11.254)                                                            \\ [-0.013cm]
11.302 & 11.302 & $   24.8\pm   10.8$ & $  13.0$ &     \ion{Fe}{18} (11.326, 11.319, 11.315)                                            \\ [-0.013cm]
11.404 & 11.406 & $   17.7\pm   21.0$ & $  18.7$ &     \ion{Fe}{22} (11.427), \ion{Fe}{18} (11.423)                                     \\ [-0.013cm]
11.476 & 11.482 & $   18.3\pm   18.2$ & $  16.7$ &     \ion{Fe}{22} (11.492, 11.505)                                                    \\ [-0.013cm]
11.517 & 11.527 & $   25.7\pm   10.7$ & $  31.9$ &     \ion{Ne}{9} (11.547)                                                             \\ [-0.013cm]
11.750 & 11.758 & $   41.0\pm   10.0$ & $  33.3$ &     \ion{Fe}{22} (11.780)                                                             \\ [-0.013cm]
11.948 & 11.947 & $   31.8\pm   11.7$ & $  27.6$ &     \ion{Fe}{21} (11.952, 11.973)                                                     \\ [-0.013cm]
12.101 & 12.106 & $   44.8\pm    8.1\phn$ & $  39.2$ &     \ion{Ne}{10} (12.132), \ion{Fe}{17} (12.124)                                      \\ [-0.013cm]
12.147 & 12.153 & $   -7.8\pm   15.1$ & $ -46.1$ &     \ion{Ne}{10} (12.132)                                                             \\ [-0.013cm]
12.251 & 12.255 & $   49.4\pm    9.9\phn$ & $  42.0$ &     \ion{Fe}{17} (12.266), \ion{Fe}{21} (12.284)                                      \\ [-0.013cm]
12.556 & 12.556 & $   37.8\pm   12.6$ & $  16.9$ &     \ion{Fe}{20} (12.576, 12.588)                                                     \\ [-0.013cm]
12.809 & 12.823 & $   60.3\pm   14.0$ & $  49.5$ &     \ion{Fe}{20} (12.846, 12.864, 12.903)                                             \\ [-0.013cm]
13.399 & 13.422 & $   57.4\pm   21.8$ & $  49.1$ &     \ion{Ne}{9} (13.447)                                                              \\ [-0.013cm]
13.495 & 13.491 & $   52.3\pm   17.1$ & $  18.2$ &     \ion{Fe}{19} (13.518, 13.497)                                                     \\ [-0.013cm]
13.567 & 13.564 & $  -18.9\pm   26.3\phn$ & $ -23.4$ &     \ion{Ne}{9} (13.553)                                                              \\ [-0.013cm]
13.694 & 13.680 & $  -44.3\pm   39.8\phn$ & $ -22.5$ &     \ion{Ne}{9} (13.699)                                                              \\ [-0.013cm]
13.777 & 13.798 & $   33.5\pm   17.1$ & $  32.6$ &     \ion{Fe}{17} (13.825)                                                             \\ [-0.013cm]
14.181 & 14.183 & $   33.5\pm   18.9$ & $  26.5$ &     \ion{Fe}{18} (14.208)                                                             \\ [-0.013cm]
14.342 & 14.348 & $   28.5\pm   39.8$ & $  21.9$ &     \ion{Fe}{18} (14.373)                                                             \\ [-0.013cm]
14.505 & 14.508 & $   37.2\pm   13.6$ & $  22.7$ &     \ion{Fe}{18} (14.534)                                                             \\ [-0.013cm]
14.622 & 14.605 & $   21.6\pm   23.4$ & $  27.1$ &     \ion{O}{8} (14.635)                                                               \\ [-0.013cm]
14.807 & 14.792 & $   18.7\pm   20.3$ & $  37.1$ &     \ion{O}{8} (14.820)                                                               \\ [-0.013cm]
14.988 & 14.987 & $   30.7\pm   24.8$ & $  53.2$ &     \ion{Fe}{17} (15.014)                                                             \\ [-0.013cm]
15.157 & 15.153 & $   22.9\pm   30.9$ & $  46.2$ &     \ion{O}{8} (15.176)                                                               \\ [-0.013cm]
15.233 & 15.236 & $   19.7\pm   18.9$ & $  33.0$ &     \ion{Fe}{17} (15.261)                                                             \\ [-0.013cm]
15.977 & 15.975 & $   47.2\pm   16.0$ & $  54.1$ &     \ion{O}{8} (16.006)                                                               \\ [-0.013cm]
17.163 & 17.168 & $   28.9\pm   15.5$ & $  39.2$  &     \ion{O}{7} (17.20)                                               \\ [-0.013cm]
17.340 & 17.367 & $   46.8\pm   19.5$ & $  39.8$  &     \ion{O}{7} (17.39)                                               \\ [-0.013cm]
17.730 & 17.733 & $   24.2\pm   32.1$ & $  56.3$ &     \ion{O}{7} (17.768)                                                               \\ [-0.013cm]
17.789 & 17.794 & $   -9.9\pm   25.5$ & $ -17.0$ &     \ion{O}{7} (17.768)                                                               \\ [-0.013cm]
18.597 & 18.593 & $   29.6\pm   13.0$ & $  65.8$ &     \ion{O}{7} (18.628)                                                               \\ [-0.013cm]
18.659 & 18.662 & $  -30.4\pm   58.6\phn$ & $ -10.7$ &     \ion{O}{7} (18.628)                                                               \\ [-0.013cm]
18.926 & 18.927 & $   60.2\pm   20.6$ & $  73.3$ &     \ion{O}{8} (18.967)                                                               \\ [-0.013cm]
18.981 & 19.001 & $  -51.1\pm   57.8\phn$ & $-244.4$ &     \ion{O}{8} (18.967)                                                               \\ [-0.013cm]
21.538 & 21.557 & $   56.1\pm   47.9$ & $  87.0$ &     \ion{O}{7} (21.602)                                                               \\ [-0.013cm]
21.594 & 21.635 & $  -68.4\pm   83.4\phn$ & $ -63.5$ &     \ion{O}{7} (21.602)                                                               \\ [-0.013cm]
21.808 & 21.814 & $  -38.1\pm  373.7$ & $-183.1$ &     \ion{O}{7} (21.807)                                                               \\ [-0.013cm]
22.113 & 22.087 & $ -161.8\pm  180.7\phn$ & $-393.8$ &     \ion{O}{7} (22.101)                                                               \\ 
\hline
\end{tabular}
\vskip 2pt
\parbox{6.1in}{ % use this to define the width of the notes under the table
\small\baselineskip 9pt
\footnotesize
\indent
$\rm ^a${A negative sign before an EW indicates an emission line.}\\
$\rm ^b${Uncertainties are $\pm 0.01$ \AA .}\\
$\rm ^c${Line not included in our model.}
}
\end{center}
\end{table*}
\normalsize

\subsection{Comparison with the Observations}
\label{comparison}

The success of the two-component model of \S~\ref{two_component} has
been evaluated in two ways. First, we compare the observed and
calculated EWs of as many lines as we can reliably measure. Here the
number of lines is about 62 compared with the 28 lines used in Paper~I.
Most of the difference is due to the iron L-shell lines now having been
computed with more reliable oscillator strengths. Second, we inspect
individual, small wavelength bands by overlaying the computed model on
the observed spectrum. This is particularly important in those regions
containing a large number of emission and absorption lines; some of
these are so crowded that there is no direct way to separate the
different contributions to the various blends (see, for example, the
line complex over the 9--17 \AA\ range in Figure~\ref{megspec} with
dozens of \ion{Fe}{17} to \ion{Fe}{24} absorption lines). Direct visual
inspection is the best way to assess the validity of the model in this
case.

The visual evaluation method is illustrated in the various panels of
Figure~\ref{megspec} where we overlay the computed spectrum (red curve)
on the observed one (black curve). This is done after a three-stage
convolution process to produce the computed line profiles. First, each
line is convolved with the microturbulence velocity representing the
local line broadening. Next, we convolve the emission lines with a
boxcar profile of 1220 \kms\ width, centered on the systemic velocity
and representing the 610 \kms\ outflow assumed for both absorption
components. We also take into account the absorption of the
emission-line photons by the line-of-sight outflowing gas which gives
the correct P~Cygni type profile. Finally, we convolve the entire
theoretical spectrum with a constant-wavelength width Gaussian
(0.023~\AA), to account for the MEG resolution used for the visual
comparison method.

The result of the comparison is quite satisfactory. The theoretical
model fits the observed spectrum over most bands and most absorption
lines. In particular, we are able to adequately fit $\sim 50$
absorption lines, including some 30 iron L-shell lines, taking into
account line blends. Most, but not all, emission lines also show a
reasonable fit. Unfortunately, most emission lines are so weak that the
quality of the fit is not, by itself, a very conclusive result.

There are several notable discrepancies deserving comment:
\begin{enumerate}
\item
Three emission lines clearly show small EWs compared with the model
predictions. These are \ion{O}{7}~(21.807, 22.101 \AA) and
\ion{O}{8}~(18.967~\AA). The discrepancy might result from low S/N and
the poor flux calibration at those wavelengths (see
\S~\ref{chandra_obs}). In fact, for two of the three lines, the
predicted EWs are within the uncertainty of the observed lines.
\item
There is a noticeable broad absorption feature at around 16--17
\AA\ that looks like a blend of many absorption lines. Sako et al.
(2001) have recently observed a similar, but much stronger, absorption
feature in the X-ray spectrum of the quasar IRAS~13349+2438 and have
identified it as an unresolved transition array of 2p-3d inner-shell
absorption lines by iron M-shell ions (\ion{Fe}{9}--\ion{Fe}{14}) in a
relatively low-ionization medium. Those authors have tentatively
associated this absorption with the ionized skin of a dusty torus
surrounding the AGN. Such absorption is not included in {\sc ion2000}.
We have verified that the abundances of these ions in the
low-ionization component are, indeed, large enough to produce the
needed extra absorption. We will make additional checks when better S/N
data become available.
\item
There is a notable discrepancy between the model and the observations
at the \ion{Si}{11} (6.813~\AA ) line. While the model
predicts absorption, the observations show an emission line at the same
energy. Both the energy and the oscillator strength of this line are
uncertain, and this might be the source of the discrepancy.
\end{enumerate} 

We have measured the absorption and emission features in the observed
spectrum using the underlying continuum from the fitted model
(\S~\ref{under_con}). Line EW measurements, together with the model
EWs, are presented in Table~\ref{lines}. In columns (1) and (2) we list
the observed wavelengths and the wavelengths measured from the model,
respectively. The observed EWs and the derived uncertainties are listed
in column (3). A negative sign before an EW indicates an emission line.
The EW uncertainties take into account the uncertainty in the continuum
placement (\S~\ref{under_con}) combined in quadrature with
uncertainties due to the number of counts in the line. These
uncertainties can be used to assess the realities of the various lines.
In column (4) we give the EW as measured from our model, and in column
(5) we identify the main ions contributing to the line or blend, based
on the model and the measured energy. The model wavelengths (column 2)
and EWs (column 4) were measured directly from the modeled spectrum
(red line in Fig~\ref{megspec}) since any attempt to calculate them is
hindered by the blending of the many components, as explained above.
The model does not include the \ion{Al}{13} (7.18~\AA) and \ion{Al}{12}
(7.76 \AA) lines which are tentatively identified in the data.

Comparing the observed and the modeled wavelengths, we find good
agreement. The mean and rms velocity shift between the two is
$-1\pm267$ \kms\ for the 67 lines ($-22\pm 245$ km\,s$^{-1}$ for the 53
absorption lines and $75\pm 338$ km\,s$^{-1}$ for the 14 emission
lines). The $\chi^2$ between the observed wavelengths and the modeled
ones is 88.7 for the 67 lines. As there is only one free parameter (the
610~\kms\ outflow velocity), we have 66 dof and the reduced $\chi^2$ is
1.32. This confirms that the adopted velocity shifts for the absorption
and emission features are in good agreement with the observation. We
checked the velocity shifts of lines which primarily originate from the
low-ionization component of our model versus lines which primarily
originate from the high-ionization component. The two ionization
components are not dynamically distinguishable within the uncertainties
of our data.

Figure~\ref{ewfig} compares the observed EWs with those derived from
the model (columns 3 and 4 of Table~\ref{lines}) except for three
low-energy emission lines (see above). The agreement between the two is
very good, both for the emission and absorption lines, as shown by the
spread along a line with slope unity. In particular, all strong lines
predicted by the model are observed. The $\chi^2$ between the observed
EWs and the modeled ones is 69.7 for the 67 lines shown. The number of
free parameters in our model is 6 (density, column density, ionization
parameter, microturbulence velocity, and global covering factors for
the two components), so the reduced $\chi^2$ is 1.14.

Our estimate for the global covering factor is based on the
emission-line intensities which, as explained earlier, are highly
uncertain (see Table~\ref{lines}). A reasonable range for the covering
factor, which is consistent with all measured EWs, is 0.2--0.7. We have
also verified visually that covering factors of 0.5 and 0.3 (for the
low-ionization and high-ionization components, respectively)
give the best fit to the emission lines at shorter wavelengths
($\lambda < 16$\AA), where the S/N is better.

The 1 keV deficit feature which was noted by George et al. (1998a) and
seems also to exist in our comparison between the {\it Chandra} data
and {\it ASCA} model (bottom panel of Figure~\ref{comparison_e}) can be
explained by many iron L-shell absorption lines around 1~keV. \ These
lines were not present in the modeling of George et al. (1998a), and
they probably cause in part the 1 keV deficit feature. George et al.
(1998a) suggested there might be two
%
%
%
%\begin{figure}
\centerline{} % just needed to provide some space.
\centerline{\includegraphics[width=8.5cm]{f8.eps}}
\figcaption{Measured EWs versus model EWs. The positive EWs are for the
absorption lines, and the negative EWs are for the emission lines. All
identified lines from Table~\ref{lines} except for 3 emission lines are
plotted here (see \S~\ref{comparison} for details). A line with a slope
of unity is drawn to guide the eye.
\label{ewfig} }
%\end{figure}
\centerline{} % just needed to provide some space.
\centerline{} % just needed to provide some space.
%
%\begin{deluxetable}{ccc}
%\tablecolumns{3}
%\tablewidth{0pt}
%\tablecaption{Model SED
%\label{powerlaw}}
\footnotesize % better fit the Journal font size.
\begin{center}
{\sc TABLE 4 \\  Model SED}
\vskip 4pt
\begin{tabular}{ccc}
\hline
\hline
{$E_1$} &
{$E_2$} &
{$\Gamma$} \\
\hline
0.2 eV          &  \phn2 eV\phm{.}   &  2.00    \\
\phm{.0}2 eV    &  40 eV\phm{.}      &  1.50    \\
\phm{.}40 eV    &  0.1 keV           &  5.77    \\
\phn0.1 keV     &  \phm{.}50 keV     &  1.77    \\
\phm{.0}50 keV  &  \phn\phn1 MeV     &  4.00    \\
\hline
\end{tabular}
\end{center}
\setcounter{table}{4}
\normalsize
\centerline{} % just needed to provide some space.
\centerline{} % just needed to provide some space.
or more zones of photoionized
material, which are changing in time, to explain this deficit. This is
also reinforced by the good fit of our two component model. We showed
that two absorption components are needed to explain the observed
richness of the absorption lines.

As mentioned in \S~\ref{Fe_L}, some \ion{Fe}{19} lines can provide
important density diagnostics, since they arise from an excited
fine-structure level with a critical density of order $10^{9}$ \cc. We
have constructed several high-density models and searched for the
strongest fine-structure lines in our data (at 13.46 and 13.96 \AA). We
found that either the lines are blended with other L-shell lines or
else the S/N is not adequate to conduct the test. Much better S/N
observations, and sole use of the HEG to keep the highest possible
resolution, are required. The current observations are inadequate.

\section{UV Absorption Toward NGC\,3783}

\subsection{The Associated UV Absorber}
\label{uv_absorber}

Many Seyfert 1 galaxies containing ionized X-ray absorbers also show
strong, narrow, UV absorption lines (e.g., Crenshaw et al. 1999 and
references therein). These features seem to vary in EW and velocity
shift on time scales of weeks to months. The suggestion that the lines
originate from the same component producing the strong X-ray absorption
(Mathur et al. 1994) has been tested in many papers (e.g., Mathur,
Elvis, \& Wilkes 1995; Shields \& Hamann 1997) without conclusive
results. The main unknowns were the SED at far-UV energies (see
Figure~\ref{SED}), responsible for the ionization of the observed UV
species, and the poor X-ray spectral resolution.

In Paper~I we compared the high-resolution {\it Chandra} spectrum of
NGC\,3783 with a UV spectrum taken $\approx 5$ yr earlier. We found
that the shifts and velocity dispersions of the X-ray absorption lines
were consistent with those observed for the UV lines. However, the
calculated warm absorber model of Paper I resulted in column densities
for \ion{C}{4} and \ion{N}{5} that were 37 and 5 times too small,
respectively, when compared with the values deduced from UV
observations. We also commented on the fact that the expected column
densities for \ion{C}{4} and \ion{N}{5} are very sensitive to the
unknown Lyman continuum SED.

We have looked again at this issue trying to resolve it in light of
newly available UV observations and the improved X-ray models. First,
we compare our {\it Chandra} spectrum with {\it HST} STIS spectra
obtained 37 days later (Crenshaw et al. 2000). The new UV observation
shows 3 absorption-line systems blueshifted with respect to the
systemic velocity by $\approx$ $-$560, $-$800, and
$-$1400~km\,s$^{-1}$. These systems have  velocity dispersions of 160,
220, and 210 km\,s$^{-1}$, respectively. Thus, the 2000 UV observation
is consistent with the {\it Chandra} line shifts, but the UV line
widths seem to be narrower than at least the resolved neon X-ray lines.
We also note that, in the 2000 UV observation, the $\approx -1400$
km\,s$^{-1}$ component became much stronger and is present also in
\ion{Si}{4}, indicating a lower ionization state compared to the other
components. It is very unlikely that such a system will also show
strong X-ray lines. Looking at Figure~\ref{vel_spec}, we see no highly
significant evidence for such absorption system in the X-ray spectrum.

We have looked at the UV lines predicted from our lower-ionization shell.
For this component we now find EW(\ion{C}{4})= 0.6 \AA\ and
EW(\ion{N}{5})=3.1 \AA, entirely consistent with the UV observations of
Crenshaw et al. (1999) and superficially confirming the idea that the
UV and X-ray lines originate from the same component. However, as
explained briefly in Paper~I, this result is extremely sensitive to the
ratio of the far-UV to 0.1--10 keV flux. The SED assumed in Paper~I
contained a much larger far-UV flux compared to the one assumed here.
As a result, the \ion{C}{4} and \ion{N}{5} absorption lines were much
weaker than for the current SED.

To verify the above conclusion, we have renormalized the UV-to-X-ray
continua in several different ways by changing the parameters of the
assumed SED from 40--200 eV (see Table~4 and
Figure~\ref{SED}). Moving only one point, at 0.1 keV, to 0.2 keV (while
keeping the X-ray slope unchanged at $\Gamma=1.77$) leads to a
reduction by a factor of 3 in EW(\ion{C}{4}) and a factor of 2.3 in
EW(\ion{N}{5}).  A further move of the point at 40 eV to 60 eV, keeping
the UV slope unchanged ($\Gamma=1.5$; see Table~4) results
in further reductions by factors of 4 and 1.4 in EW(\ion{C}{4}) and
EW(\ion{N}{5}), respectively. Both changes do not influence the chosen
value of $U_{\rm oxygen}$ and have no effect on the calculated X-ray
line EWs. Thus the EWs of the UV absorption lines are extremely
sensitive to the exact cut-off frequency of the unobserved UV
continuum. They are also sensitive, by a lesser amount, to the relative
UV-to-X-ray normalization. The conclusion is that there is little one
can infer from the comparison of the available UV and X-ray
observations regarding the relationship between the UV and X-ray
absorbers. Part of the solution is to obtain real simultaneous
observations. These will give the simultaneous EWs for the X-ray and UV
lines and some handle on the relative UV and X-ray continuum fluxes.

\subsection{The High-Velocity Cloud Toward NGC\,3783}

Toward the direction of NGC\,3783 there is a high-velocity cloud (HVC)
denoted HVC 287.5+22.5+240 (Lu et al. 1998 and references therein).
HVCs are \ion{H}{1} clouds moving at velocities inconsistent with
simple models of Galactic rotation and are generally defined as having
velocities exceeding 90 km\,s$^{-1}$. The HVC toward NGC\,3783 has a
velocity of 240 km\,s$^{-1}$ with respect to the local standard of rest
and is at a distance of 10--50 kpc. In addition to \ion{H}{1} absorption it was
found to have absorption by \ion{S}{2} and \ion{Fe}{2}. We find it
unlikely that the absorption we report on in this paper has
anything to do with this HVC for several reasons: (1) the velocity of
this HVC is an order of magnitude less than the redshift velocity of
NGC\,3783, corresponding to $\approx 4$ resolution elements of the MEG
at 10 \AA , (2) the column densities derived for this HVC
[$N$(\ion{H}{1}) = $(8\pm 1)\times 10^{19}$ cm$^{-2}$, $N$(\ion{Fe}{2})
= $(8.5\pm 1.0)\times 10^{13}$ cm$^{-2}$, and $N$(\ion{S}{2}) =
$(3.7\pm 1.0)\times 10^{14}$ cm$^{-2}$] are about two orders of
magnitude smaller than the column densities indicated from the
NGC\,3783 spectrum, and (3) in the HVC there is no source of radiation
to cause the high degree of ionization needed to produce the ions
detected in the NGC\,3783 spectrum.

\section{The Iron K$\alpha$ Line}
\label{feka}

At least some of the Fe\,K$\alpha$ fluorescent line emission seen from
Seyfert 1 galaxies is often very broad and is generally interpreted as
originating from an accretion disk orbiting around a central black hole
(e.g., Fabian et al. 2000 and references therein). The profile of the
line is assumed to reflect gravitational and Doppler effects within a
few tens of gravitational radii of the black hole. As narrow
Fe\,K$\alpha$ line emission is also expected from matter beyond the
accretion disk, it is important to separate the narrow emission from
the broad emission. The spectral resolution of pre-{\it Chandra} X-ray
missions was not high enough to resolve the narrow component from the
broad one (e.g., the $\sim$ 160 eV FWHM of {\it ASCA} is too poor for
that task), and currently only the {\it Chandra} HETGS is capable of
such separation (with a factor of $\sim 4$ better spectral resolution
at Fe\,K$\alpha$ than {\it ASCA}).

\begin{figure*}
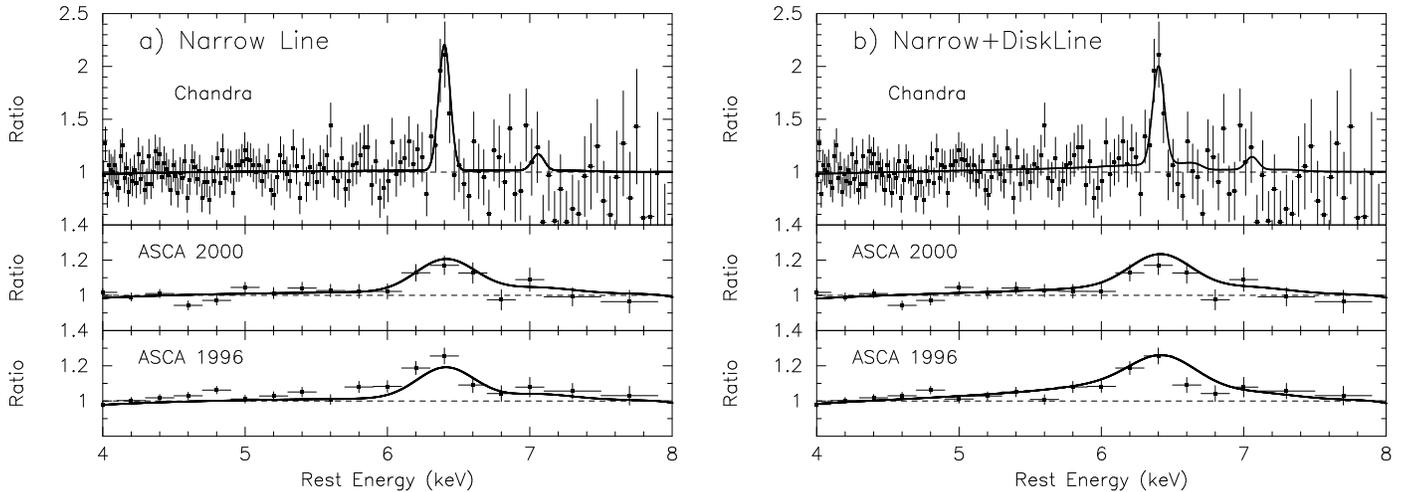

\hglue0.0cm{\includegraphics[angle=-90,width=8.9cm]{f9a.eps}}
\hglue0.5cm{\includegraphics[angle=-90,width=8.9cm]{f9b.eps}}
\caption{ The observed ratios of the data to the underlying continua
near the Fe\,K$\alpha$ line, along with model predictions (bold line;
convolved with the appropriate instrumental response) discussed in the
text. For {\it ASCA}, the mean ratio is shown for the SIS detectors for
all observations at each epoch, rebinned for clarity. Fe K$\beta$ is
included at 7.06~keV with a flux that is 11\% that of Fe K$\alpha$. a)
The model prediction for a narrow Gaussian Fe K$\alpha$ line at 6.4~keV
and a flux of $6.6\times 10^{-5}$ photons\,cm$^{-2}$\,s$^{-1}$ as
described in \S~\ref{feka-nl}. Such a model provides an adequate
description of the {\it Chandra} and 2000 {\it ASCA} data sets.
However, the 1996 {\it ASCA} data show an excess in the red wing of the
line. b) The model prediction for a narrow Gaussian plus a 6.4~keV
Schwarzschild ``diskline'' where the disk is inclined at an angle
$i=30\degr$ and the line emissivity $\propto r^{-3}$ between 6 and
$10^3$ $r_{\rm g}$ (see \S~\ref{feka-bl}). In the upper two panels the
narrow component has a flux of $6.6\times 10^{-5}$
photons\,cm$^{-2}$\,s$^{-1}$ and the broad component $5.0\times
10^{-5}$ photons\,cm$^{-2}$\,s$^{-1}$ (the maximum intensity consistent
with the data). In the lower panel, the narrow and broad components
have fluxes $4.2$ and $12\times 10^{-5}$ photons\,cm$^{-2}$\,s$^{-1}$,
respectively (see \S~\ref{feka-bl}).
\label{fe} }
\end{figure*}

\subsection{The Narrow Component}
\label{feka-nl}

The narrow Fe\,K$\alpha$ line from NGC\,3783 is very prominent in the
{\it Chandra} HETGS spectrum (Figure~\ref{fe}a). The line peak is at
$6393\pm15$ eV, consistent with Fe\,{\sc i}--{\sc xii}. Modeling the
line in the combined spectrum as a single Gaussian, we find a FWHM of
$26\pm 10$ m\AA\ which is consistent with the line not being resolved
at the resolution of the MEG. The line flux is $(6.6\pm 2.1)\times
10^{-5}$ photons cm$^{-2}$ s$^{-1}$ $=$ $(6.8\pm 2.2)\times 10^{-13}$
ergs cm$^{-2}$ s$^{-1}$, corresponding to an EW of 35$\pm$11
m\AA\ ($115\pm36$ eV) as measured against the underlying continuum.
Investigating the HEG spectrum alone, the FWHM of the line is $15\pm6$
m\AA . This FWHM is consistent with the HEG resolution of 12 m\AA\, and
thus the line has a FWHM of less than 3250~\kms. This velocity is
comparable to, or lower than, the velocity of the BLR of this object
which is $\sim$ 4000 km\,s$^{-1}$ (Reichert et al. 1994; Wandel,
Peterson, \& Malkan 1999; FWHM$_{\rm H\beta} = 4100\pm 1160$ \kms).
Given this upper limit on the narrow Fe K$\alpha$ width, and assuming a
simple correlation of line width with radial location in a Keplerian
system, we conclude that the origin of this feature is anywhere from
just outside the BLR to as far as the NLR. In particular, the observed
energy and large EW suggest that the line may arise in the putative
torus.

Ghisellini, Haardt, \& Matt (1994) and Krolik, Madau, \& Zycki (1994)
have calculated the X-ray line emission expected from putative
obscuring tori in AGNs. Both studies predict the Fe~K$\alpha$ line to
have a characteristic EW of $\sim 100$ eV for a torus which has an
opening angle of $\sim 35\degr$, an inclination angle to the line of
sight of $\la 35\degr$, and a column density $N_{\rm H} \ga 10^{24}$
cm$^{-2}$. Later, Matt, Brandt, \& Fabian (1996) considered changes in
the line EW for various metalicities. All predicted EWs are in
agreement with the one measured here and suggest that the narrow
Fe~K$\alpha$ line from NGC\,3783 could arise from the torus.

In a recent paper, Yaqoob et al. (2001) report the detection of a
narrow Fe K$\alpha$ line from NGC\,5548. In Paper~I we indicated the
resemblance of the high-resolution X-ray spectrum of NGC\,3783 to the
one of NGC\,5548 regarding the emission and absorption features at
lower energies (Kaastra et al. 2000). Also in the narrow Fe K$\alpha$
line properties we find a good resemblance. In both objects the peak
energy of the line is consistent with the rest energy for neutral iron
(6.4~keV), and the lines' widths, EWs, and intensities are comparable.
In NGC\,5548 the narrow Fe K$\alpha$ line is resolved (${\rm FWHM} =
4515^{+3525}_{-2645}$) and is consistent with being produced in the
outer BLR; this FWHM is also consistent with the upper limit we derived
for NGC\,3783.  Yaqoob et al. (2001) do not detect the broad
Fe\,K$\alpha$ line in the {\it Chandra} HETGS spectrum of NGC\,5548 due
to the small effective area of the instrument, but they find the
statistical upper limits to be consistent with earlier {\it ASCA}
measurements. The situation is similar for NGC\,3783 as explained
below.

\subsection{The Broad Component}
\label{feka-bl}

A narrow Gaussian emission line does not provide an adequate
description of the Fe\,K$\alpha$ emission observed in a number of
previous {\it ASCA} spectra of NGC\,3783. For instance, George et al.
(1998a) found the line to be asymmetric and skewed to lower energies
during the 1993 and 1996 observations. Our new {\it ASCA\/} data
confirm this result. In the lower panels of Figure~\ref{fe}a we show
that while the narrow line derived above is consistent with the 2000
{\it ASCA} data, the 1996 {\it ASCA} data show an excess in the red
wing of the line. Here we investigate the constraints that can be
placed on an accretion-disk line component in the light of our {\it
Chandra} results. The broad component is not detected in the {\it
Chandra} data, and we only derive an upper limit for it.

Following George et al. (1998a), in this section our model for the Fe
K$\alpha$ emission consists of a narrow Gaussian (energy and width
constrained as found above) plus a ``diskline'' component for a
Schwarzschild black hole (see Fabian et al. 1989). The latter
represents the profile expected from the innermost regions of the
accretion flow. It is parameterized assuming a plane-parallel geometry
inclined at an angle, $i$, to the line of sight, and a line emissivity
as a function of radius, $r$, proportional to $r^{-q}$ over the range
$R_{\rm i} \leq r \leq R_{\rm o}$. There is a degeneracy in such a
model since for $q < 2$ and $R_{\rm i} \gg 6r_{\rm g}$ (where $r_{\rm
g}$ is the gravitational radius of the black hole) the relativistic and
Doppler broadening of the diskline is negligible and hence the profile
is approximately Gaussian. Thus here we limit $2 \leq q \leq 3$,
$6r_{\rm g} \leq R_{\rm i} \leq 60r_{\rm g}$ and fix $R_{\rm o} =
10^3r_{\rm g}$. Fe K$\beta$ emission was included at an intensity 11\%
of that of the corresponding K$\alpha$ component.

We modeled the {\it Chandra} data assuming the underlying continuum
derived in \S4.1, using the simultaneous {\it RXTE} data  to constrain
the strength of the reflection component and allowing all parameters in
the model to be free.  The {\it Chandra\/} data give 90\% confidence
limits on the intensities of the lines of $A_{\rm nl} \lesssim 1.0
\times 10^{-4}\ {\rm photons\ cm^{-2}\ s^{-1}}$ and $A_{\rm bl}
\lesssim 1.6\times10^{-4}\ {\rm photons\ cm^{-2}\ s^{-1}}$ for the
narrow and broad components, respectively. Unfortunately, the {\it
Chandra} data do not allow us to constrain simultaneously $q$, $R_{\rm
i}$, $i$, and the rest-frame energy ($E_{\rm bl}$) of the diskline.
Even when the {\it Chandra} and {\it ASCA} data are combined these
parameters are not constrained.  Thus we have further restricted our
analysis to an ``extreme'' diskline with $q=3$, $R_{\rm i}=6r_{\rm g}$,
$E_{\rm bl}=6.4$~keV and $i=30\degr$. We find such a model to be
consistent with the 2000 {\it Chandra} and {\it ASCA} data sets when
$A_{\rm bl} \lesssim 5\times10^{-5}\ {\rm photons\ cm^{-2}\ s^{-1}}$
(EW $\lesssim 70$~eV). The narrow-line component is consistent with
those found above [$A_{\rm nl} = (6.2\pm2.0)\times10^{-5}\ {\rm
photons\ cm^{-2}\ s^{-1}}$; EW $=100\pm30$~eV]. The profiles from such
a model are compared to the {\it Chandra} and {\it ASCA} data in the
upper two panels of Figure~\ref{fe}b.

We have performed a re-analysis of the Fe\,K$\alpha$ line emission in
the earlier {\it ASCA} data sets. Again all the parameters associated
with a general diskline cannot be constrained simultaneously by the
{\it ASCA} data. However for an ``extreme'' diskline, as discussed
above, we find $A_{\rm bl} = (12\pm4)\times10^{-5}\ {\rm
photons\ cm^{-2}\ s^{-1}}$ (EW $=160\pm60$~eV) and $A_{\rm nl} =
(4.2\pm1.6)\times10^{-5}\ {\rm photons\ cm^{-2}\ s^{-1}}$ (EW
$=50\pm20$~eV) for the combined 1996 data. This profile is shown in the
lower panel of Figure~\ref{fe}b.  We note that the EW of the broad
component is somewhat lower than the $230\pm60$~eV reported by George
et al. (1998a). This is primarily due to our inclusion of the
Compton-reflection component.

We conclude that both the intensity and the profile of the
Fe\,K$\alpha$ line emission changed between 1996 and 2000.
Specifically, we find evidence that the EW of the broad component of
the line decreased between the observations made at these epochs.
Changes in the profile, intensity and/or EW of the broad component have
been seen in a number of other AGN (e.g., Fabian et al. 2000 and
references therein). Unfortunately, the large number of free parameters
(associated with both the Fe\,K$\alpha$ line emission and the reflected
continuum) and the quality of the data currently available prevent us
placing stringent constraints on the cause of such variability in
NGC\,3783.

\section{Summary}

The high-resolution X-ray spectrum of NGC\,3783 shows several dozen
absorption lines and a few emission lines from the H-like and He-like
ions of O, Ne, Mg, Si, and S as well as from \ion{Fe}{17}--\ion{Fe}{23}
L-shell transitions. We reanalyzed the spectrum (first presented in Paper~I)
using better flux and wavelength calibrations. Combining several lines
from each element we demonstrated the existence of the absorption lines
and determined they are blueshifted relative to the systemic velocity
by $-610\pm 130$ \kms . For the Ne lines in the HEG spectrum, we find
${\rm FWHM}=840^{+490}_{-360}$~km\,s$^{-1}$; no other lines
are resolved.

We used LFZs to determine the X-ray continuum (which yielded $\Gamma =
1.77^{+0.10}_{-0.11}$) and photoionization modeling (assuming
thermal and photoionization equilibrium) to model the absorption and
emission lines. We used a model which includes two
absorption components that differ by an order of magnitude in their
ionization parameters. The two components are spherically outflowing
from the AGN and thus contribute, via P~Cygni profiles, to both the
absorption as well as the emission lines. In our fitted model the
low-ionization component has $\log U_{\rm oxygen}=-1.745$ and a global
covering factor of 0.5. The high-ionization component has $\log U_{\rm
oxygen}=-0.745$ and a global covering factor of 0.3. Both components
have line-of-sight covering factors of unity, hydrogen column densities
of $\log N_{\rm H}=22.2$, gas densities of 10$^8$ cm$^{-3}$, and
microturbulence velocities of 300~\kms. The line EWs calculated from
the model agree well with the EWs measured from the data. Since there
are many blended lines in the spectrum, and the quality of the data
does not enable deblending, we also compared the model to the data by
overploting them. This comparison further demonstrates the good agreement
between the data and our model.

The rich X-ray absorption spectrum of NGC\,3783 suggests a complex
model for this object in particular and for AGNs with warm absorbers in
general. The model emerging for the X-ray absorption includes two
absorbing components, probably related to those seen in UV studies
(e.g., Crenshaw et al. 1999; M. J. Collinge et al., in preparation)
where several UV absorption systems are resolved. There may well be
several X-ray absorbing systems present, but with the current data
quality we can only model two components. We find the low-ionization
component of our model can plausibly produce UV absorption lines with
the EWs observed. However, this result is highly sensitive to the
unobservable UV-to-X-ray continuum. Although the outflow velocity of
the X-ray absorber in NGC\,3783 is consistent with that of the UV
absorber, the low S/N of the high-resolution X-ray spectrum does not
allow conclusive comparison. This must await longer observations
simultaneous with the UV.

We find the X-ray data from the {\it Chandra} observation to agree well
with simultaneous observations from {\it ASCA} and {\it RXTE\/} both in
spectral shape and in the time variability of $\approx 30\%$ over a few
hours. The inconsistencies in absolute fluxes between the three
observations are consistent with the known uncertainties of the
instruments. Using the {\it ASCA} observations we demonstrated that
during the {\it Chandra} observation NGC\,3783 was in a typical state.
We are also able to explain the 1 keV deficit found in previous
modeling of {\it ASCA} data; this feature is probably arising from the
many iron L-shell lines not considered previously in the models that
are now detected in the high-resolution X-ray spectrum.

We have found a prominent narrow Fe~K$\alpha$ emission line which is
not resolved. We set an upper limit on its FWHM of 3250~\kms. This is
consistent with this line originating outside the BLR. The line flux is
consistent with predictions from models of torus emission. The data
cannot effectively constrain any broad Fe~K$\alpha$ emission though
they are consistent within the uncertainties with the broad line found
previously from this object.

Comparing Figures~\ref{megspec} and~\ref{comparison_e} clearly
demonstrates the increase in our understanding of NGC\,3783 when
advancing from {\it ASCA} resolution to {\it Chandra} HETGS resolution.
As part of a large collaboration we have been allocated 850 ks to
further observe this object with the {\it Chandra} HETGS to monitor the
X-ray continuum and line variations. With the help of these new
observations, we will be able to study further and resolve the X-ray
spectrum of NGC\,3783. Together with simultaneous {\it HST} STIS
observations, this campaign should enable us to determine the nature,
origin, and evolution of the X-ray and UV absorbers by a direct
comparison of their dynamical and physical properties.

\acknowledgments
We thank all the members of the {\it Chandra} team for their enormous
efforts. We thank H. L. Marshall for helpful discussions.  We are
grateful for several valuable suggestions by F. W. Hamann.  We
gratefully acknowledge the financial support of NASA grant NAS~8-38252
(G. P. G., PI), NASA LTSA grant NAG~5-8107 (S. K., W. N. B.), NASA
grant NAG~5-7282 (S. K., W. N. B.), the Alfred P. Sloan Foundation (W.
N.  B.), and the Israel Science Foundation and the Jack Adler Chair for
Extragalactic Astronomy (H. N.).

%\clearpage


\begin{thebibliography}{}

\bibitem[Alloin et al.\ (1995)]{1995A&A...293..293A} Alloin, D. et al.
1995, \aap, 293, 293

\bibitem[Arnaud(1996)]{1996adass...5...17A} Arnaud, K. A. 1996, in
Astronomical Data Analysis Software and Systems V, ed. Jacoby G. and
Banes J. (San Francisco: ASP) p. 17

\bibitem[Bar-Shalom 1998]{hullac98} Bar-Shalom, A., Klapisch, M., \&
Oreg, J. 2001, J. Quant. Spectr. Radiat. Transfer, in press

\bibitem[Behar et al. 2001]{behar00} Behar, E., Cottam, J., \& Kahn, S.
M. 2001, ApJ, 548, in press (astro-ph/0003099)

\bibitem[Bevington \& Robinson]{B92} Bevington, P. R., \& Robinson, D.
K. 1992, Data Reduction and Error Analysis for the Physical Sciences:
Second Edition. McGraw Hill, New York

\bibitem[Bradt, Rothschild and Swank (1993)]{1993A&AS...97..355B} Bradt, H. 
V., Rothschild, R. E., \& Swank, J. H. 1993, \aaps, 97, 355 

\bibitem[Brandt, Mathur, Reynolds \& Elvis(1997)]{1997MNRAS.292..407B} 
Brandt, W. N., Mathur, S., Reynolds, C. S., \& Elvis, M. 1997, \mnras, 
292, 407 

\bibitem[Brandt \& Schulz (2000)]{B2000} Brandt, W. N., \& Schulz, N. S.
2000, \apjl, 544, L123

\bibitem[Crenshaw et al. (1999)]{1999ApJ...516..750C} Crenshaw, D. M. , 
Kraemer, S. B., Boggess, A. , Maran, S. P., Mushotzky, R. F., \& Wu, C. -C. 
1999, \apj, 516, 750 

\bibitem[Crenshaw, Kraemer and Ruiz (2000)]{2000AAS...196.1607C}
Crenshaw, D. M., Kraemer, S. B., \& Ruiz, J. R. 2000, \baas, 32, 695

\bibitem[de Vaucouleurs et al.(1991)]{1991trcb.book.....D} de
Vaucouleurs, G., de Vaucouleurs, A., Corwin, H. G., Buta, R. J.,
Paturel, G., \& Fouque, P. 1991, Third Reference Catalogue of Bright
Galaxies (Springer-Verlag: New York)

\bibitem[Fabian, Rees, Stella \& White(1989)]{1989MNRAS.238..729F} Fabian, 
A. C., Rees, M. J., Stella, L., \& White, N. E. 1989, \mnras, 238, 729 

\bibitem[Fabian et al.(1994)]{1994PASJ...46L..59F} Fabian, A. C., et al. 
1994, \pasj, 46, L59 

\bibitem[Fabian et al. (2000)]{2000PASP..112.1145F} Fabian, A. C.,
Iwasawa, K., Reynolds, C. S., \& Young, A. J. 2000, \pasp, 112, 1145

\bibitem[Gehrels (1986)]{1986ApJ...303..336G} Gehrels, N. 1986, \apj, 303, 336 

\bibitem[George (1995)]{G95} George, I., M., Turner, T., J., \& Netzer,
H. 1995, \apj, 438, L67

\bibitem[George et al. (1998a)]{1998ApJ...503..174G} George, I. M., Turner, T. 
J., Mushotzky, R., Nandra, K., \& Netzer, H. 1998a, \apj, 503, 174 

\bibitem[George et al. (1998b)]{1998ApJS..114...73G} George, I. M.,
Turner, T. J., Netzer, H. , Nandra, K., Mushotzky, R. F., \& Yaqoob,
T. 1998b, \apjs, 114, 73

\bibitem[George et al.(2000)]{2000ApJ...531...52G} George, I. M., Turner, 
T. J., Yaqoob, T., Netzer, H., Laor, A., Mushotzky, R. F., Nandra, K., \& 
Takahashi, T. 2000, \apj, 531, 52 

\bibitem[Ghisellini, Haardt \& Matt(1994)]{1994MNRAS.267..743G} Ghisellini, 
G., Haardt, F., \& Matt, G. 1994, \mnras, 267, 743 

\bibitem[Halpern(1984)]{1984ApJ...281...90H} Halpern, J. P. 1984, \apj, 
281, 90 

\bibitem[Kaastra et~al. (2000)]{astro-ph..0002345} Kaastra, J. S., Mewe, R.,
Liedahl, D. A., Komossa, S., \& Brinkman, A. C. 2000, A\&A, 354, L83

\bibitem[Kaspi et al. (2000)]{k2000} Kaspi, S., Brandt, W. N., Netzer, H., 
Sambruna, R., Chartas, G., Garmire, G. P., \& Nousek, J. A. 2000, \apj, 535, L17
(Paper~I)

\bibitem[Klapisch et al. (1977)]{1977OSAJ...67..148K} Klapisch, M.,
Schwob, J. L., Fraenkel, B. S., \& Oreg, J. 1977, J. Opt. Soc. A., 67, 148
%
% Journal of the Optical Society of America

\bibitem[Krolik & Kriss (1995)]{1995ApJ...447..512K} Krolik, J. H., \& Kriss, 
G. A. 1995, \apj, 447, 512 (erratum 456, 909 [1996])

\bibitem[Krolik, Madau \& Zycki(1994)]{1994ApJ...420L..57K} Krolik, J. H., 
Madau, P., \& Zycki, P. T. 1994, \apjl, 420, L57 

\bibitem[Lu et al.(1998)]{1998AJ....115..162L} Lu, L., Sargent, W. L. W., 
Savage, B. D., Wakker, B. P., Sembach, K. R., \& Oosterloo, T. A. 
1998, \aj, 115, 162 

\bibitem[Magdziarz and Zdziarski (1995)]{1995MNRAS.273..837M} Magdziarz, P., \&
Zdziarski, A. A. 1995, \mnras, 273, 837 

\bibitem[Maran et al.(1996)]{1996ApJ...465..733M} Maran, S. P., Crenshaw, 
D.\ M., Mushotzky, R. F., Reichert, G. A., Carpenter, K. G., Smith, A.
M., Hutchings, J. B., \& Weymann, R. J. 1996, \apj, 465, 733 

\bibitem[Mathur, Elvis \& Wilkes (1995)]{1995ApJ...452..230M} Mathur, S., 
Elvis, M., \& Wilkes, B. 1995, \apj, 452, 230 

\bibitem[Mathur, Wilkes, Elvis \& Fiore(1994)]{1994ApJ...434..493M} Mathur, 
S., Wilkes, B., Elvis, M., \& Fiore, F. 1994, \apj, 434, 493 

\bibitem[Matt, Brandt \& Fabian(1996)]{1996MNRAS.280..823M} Matt, G., 
Brandt, W. N., \& Fabian, A. C. 1996, \mnras, 280, 823 

\bibitem[Netzer (1993)]{1993ApJ...411..594N} Netzer, H. 1993, \apj, 411, 594 

\bibitem[Netzer (1996)]{1996ApJ...473..781N} Netzer, H. 1996, \apj, 473, 781 

\bibitem[Netzer, Turner \& George(1998)]{1998ApJ...504..680N} Netzer, H., 
Turner, T. J., \& George, I. M. 1998, \apj, 504, 680 

\bibitem[Nicastro Fiore \& Matt (1999)]{1999ApJ...517..108N} Nicastro, F., 
Fiore, F., \& Matt, G. 1999, \apj, 517, 108

\bibitem[Otani et al.(1996)]{1996PASJ...48..211O} Otani, C., et al. 1996, 
\pasj, 48, 211 

\bibitem[Reichert et al.\ (1994)]{1994ApJ...425..582R} Reichert, G. A. 
et al. 1994, \apj, 425, 582 

\bibitem[Reynolds(1997)]{1997MNRAS.286..513R} Reynolds, C. S. 1997, 
\mnras, 286, 513 

\bibitem[Reynolds \& Fabian (1995)]{1995MNRAS.273.1167R} Reynolds, C.
S., \& Fabian, A. C. 1995, \mnras, 273, 1167

\bibitem[Sako, Kahn, Paerels \& Liedahl(2000)]{2000ApJ...543L.115S} Sako, 
M., Kahn, S. M., Paerels, F., \& Liedahl, D. A. 2000, \apjl, 543, L115 

\bibitem[Sako et al.(2001)]{2001A&A...365L.168S} Sako, M., et al. 2001, 
\aap, 365, L168 

\bibitem[Shields \& Hamann (1997)]{1997ApJ...481..752S} Shields, J. C., \& 
Hamann, F. 1997, \apj, 481, 752 

\bibitem[Stark et al. (1992)]{1992ApJS...79...77S} Stark, A. A., Gammie, 
C. F., Wilson, R. W., Bally, J., Linke, R. A., Heiles, C., \& Hurwitz, 
M. 1992, \apjs, 79, 77

\bibitem[Tanaka, Inoue and Holt (1994)]{1994PASJ...46L..37T} Tanaka, Y., 
Inoue, H., \& Holt, S. S. 1994, \pasj, 46, L37 

\bibitem[Turner et al.(1993)]{1993ApJ...419..127T} Turner, T. J., Nandra, 
K., George, I. M., Fabian, A. C., \& Pounds, K. A. 1993, \apj, 419, 127 

\bibitem[Ulvestad \& Wilson(1984)]{1984ApJ...285..439U} Ulvestad, J. S., 
\& Wilson, A. S. 1984, \apj, 285, 439 

\bibitem[Wandel, Peterson and Malkan (1999)]{1999ApJ...526..579W} Wandel, 
A., Peterson, B. M., \& Malkan, M. A. 1999, \apj, 526, 579 

\bibitem[Wise 1999]{W1999} Wise, M. W., Davis, J. E., Huenemoerder, 
D. P., Houck, J. C., \& Dewey, D. 2000, MARX 3.0 Technical Manual (unpublished)

\bibitem [Yaqoob (2000a)]{Y2000a} Yaqoob, T. 2000a, {\it ASCA} GOF
Calibration Memo ASCA-CAL-00-06-01 (v1.0)

\bibitem [Yaqoob (2000b)]{Y2000b} Yaqoob, T. 2000b, {\it ASCA} GOF
Calibration Memo ASCA-CAL-00-06-02 (v1.0)

\bibitem[Yaqoob et al.(2001)]{2001ApJ...546..759Y} Yaqoob, T., George, I. 
M., Nandra, K., Turner, T. J., Serlemitsos, P. J., \& Mushotzky, R. F. 
2001, \apj, 546, 759 

\bibitem[Yaqoob \& Serlemitsos(2000)]{2000ApJ...544L..95Y} Yaqoob, T., \& 
Serlemitsos, P. 2000, \apjl, 544, L95 

\end{thebibliography}
\end{document}